\documentclass[conference]{IEEEtran}
\usepackage{cite}
\usepackage{amsmath,amssymb,amsfonts}
\usepackage{algorithmic}
\usepackage{graphicx}
\usepackage{textcomp}
\usepackage{xcolor}
\usepackage{fancyhdr}
\usepackage[hyphens]{url}
\usepackage[normalem]{ulem}
\usepackage[final]{microtype}
\usepackage[keeplastbox]{flushend}
\usepackage{amsmath,amsfonts,amsthm,amssymb}
\usepackage{subcaption}
\usepackage{multirow}
\usepackage{color,soul}
\usepackage{listings}             
\usepackage[compact]{titlesec}
\usepackage{paralist}
\usepackage{multirow, makecell}
 \usepackage{lscape}

\newcommand{\ec}[1]{{{\color{blue}EC: {#1}}}}

\titlespacing{\section}{0pt}{1ex}{0.5ex}
\titlespacing{\subsection}{0pt}{0.75ex}{0ex}
\titlespacing{\subsubsection}{0pt}{0.5ex}{0ex}

\def\BibTeX{{\rm B\kern-.05em{\sc i\kern-.025em b}\kern-.08em
    T\kern-.1667em\lower.7ex\hbox{E}\kern-.125emX}}

\renewcommand{\baselinestretch}{0.97}

\DeclareRobustCommand{\hlcyan}[1]{{\sethlcolor{white}\hl{#1}}}
\DeclareRobustCommand{\shepherd}[1]{{\sethlcolor{white}\hl{#1}}}

\sethlcolor{white}

\fancypagestyle{firstpage}{
  \fancyhf{}

}

\title{Automatic Microprocessor Performance Bug Detection} 

\author{
\IEEEauthorblockN{
Erick~Carvajal~Barboza\IEEEauthorrefmark{1},
Sara~Jacob\IEEEauthorrefmark{1},
Mahesh~Ketkar\IEEEauthorrefmark{2},
Michael~Kishinevsky\IEEEauthorrefmark{2},
Paul~Gratz\IEEEauthorrefmark{1},
Jiang~Hu\IEEEauthorrefmark{1}
}
\IEEEauthorblockA{
\IEEEauthorrefmark{1}Texas A\&M University\\
\{ecarvajal, sarajacob\_96, jianghu\}@tamu.edu, pgratz@gratz1.com
}
\IEEEauthorrefmark{2}Intel Corportation\\
\{mahesh.c.ketkar, michael.kishinevsky\}@intel.com
}

\begin{document}
\maketitle
\thispagestyle{firstpage}
\pagestyle{plain}

\begin{abstract}
Processor design validation and debug is a difficult and complex task,
which consumes the lion's share of the design process.  Design bugs
that affect processor performance rather than its functionality are
especially difficult to catch, par\-ti\-cu\-lar\-ly in new
microarchitectures.  This is because, unlike functional bugs, the
correct processor performance of new microarchitectures on complex,
long-running benchmarks is typically not deterministically known.
Thus, when performance benchmarking new microarchitectures,
performance teams may assume that the design is correct when the
performance of the new microarchitecture exceeds that of the previous
generation, despite significant performance re\-gres\-sions existing
in the design.  In this work we present a two-stage, machine
learning-based methodology that is able to detect the existence of
performance bugs in microprocessors.  \hlcyan{Our results show that
our best technique detects 91.5\% of microprocessor core performance
bugs whose average IPC impact across the studied applications is
greater than 1\% versus a bug-free design with zero false
positives. When evaluated on memory system bugs, our technique
achieves 100\% detection with zero false positives.}  Moreover, the
detection is automatic, requiring very little performance engineer
time.

\end{abstract}

\section{Introduction}
\label{sec:intro}



Verification and validation are typically the largest component of the
design effort of a new processor. The effort can be broadly divided
into two distinct disciplines, namely, functional and performance
verification. The former is concerned with design correctness and has
rightfully received significant attention in the literature. Even
though challenging, it benefits from the availability of known correct
output to compare against. Alternately, performance verification,
which is typically concerned with generation-over-generation workload
performance improvement, suffers from the lack of a known correct
output to check against.  Given the complexity of computer systems, it
is often extremely difficult to accurately predict the expected
performance of a given design on a given workload.  Performance bugs,
nevertheless, are a critical concern as the cadence of process
technology scaling slows, as they rob processor designs of the primary
performance and efficiency gains to be had through improved
microarchitecture.

Further, while simulation studies may provide an approximation of how
long a given application might run on a real system, \hl{prior work
  has shown that these performance estimates will typically vary
  greatly from real system
  performance}~\cite{Black1996,simalpha,harmfulsim}.  \hl{Thus,
  the 
  process to detect performance bugs, in practice, is via tedious and
  complex manual analysis, often taking large periods of time to
  isolate and eliminate a single bug.}  For example, consider the case
of the Xeon Platinum 8160 processor snoop filter eviction
bug~\cite{mccalpin2018hpl}.  Here the performance regression was
$>10\%$ on several benchmarks.  The bug took several months of
debugging effort before being fully characterized.  In a competitive
environment where time to market \emph{and} performance are essential,
there is a critical need for new, more automated mechanisms for
performance debugging.





\hl{To date, little prior research in performance debugging exists,
  despite its importance and difficulty. The few existing
  works}~\cite{Bose1994,surya1994architectural,Singhal2004} \hl{take
  an approach of manually constructing a reference model for
  comparison.  However, as described by Ho \emph{et
    al.}}~\cite{Ho1995}, \hl{the same bug can be repeated in the model
  and thus cannot be detected.  Moreover, building an effective model
  is very time consuming and labor intensive.}  \hl{ Thus, as a first
  step we focus on automating bug detection.  We note that, simply
  detecting a bug in the pre-silicon phase or early in the post-silicon phase 
  is highly valuable to shorten the validation process or to avoid costly steppings.
  Further, bug detection is a challenge by itself.}

\begin{figure}[!hbt]
    \centering
     \includegraphics[width = 0.45\textwidth]{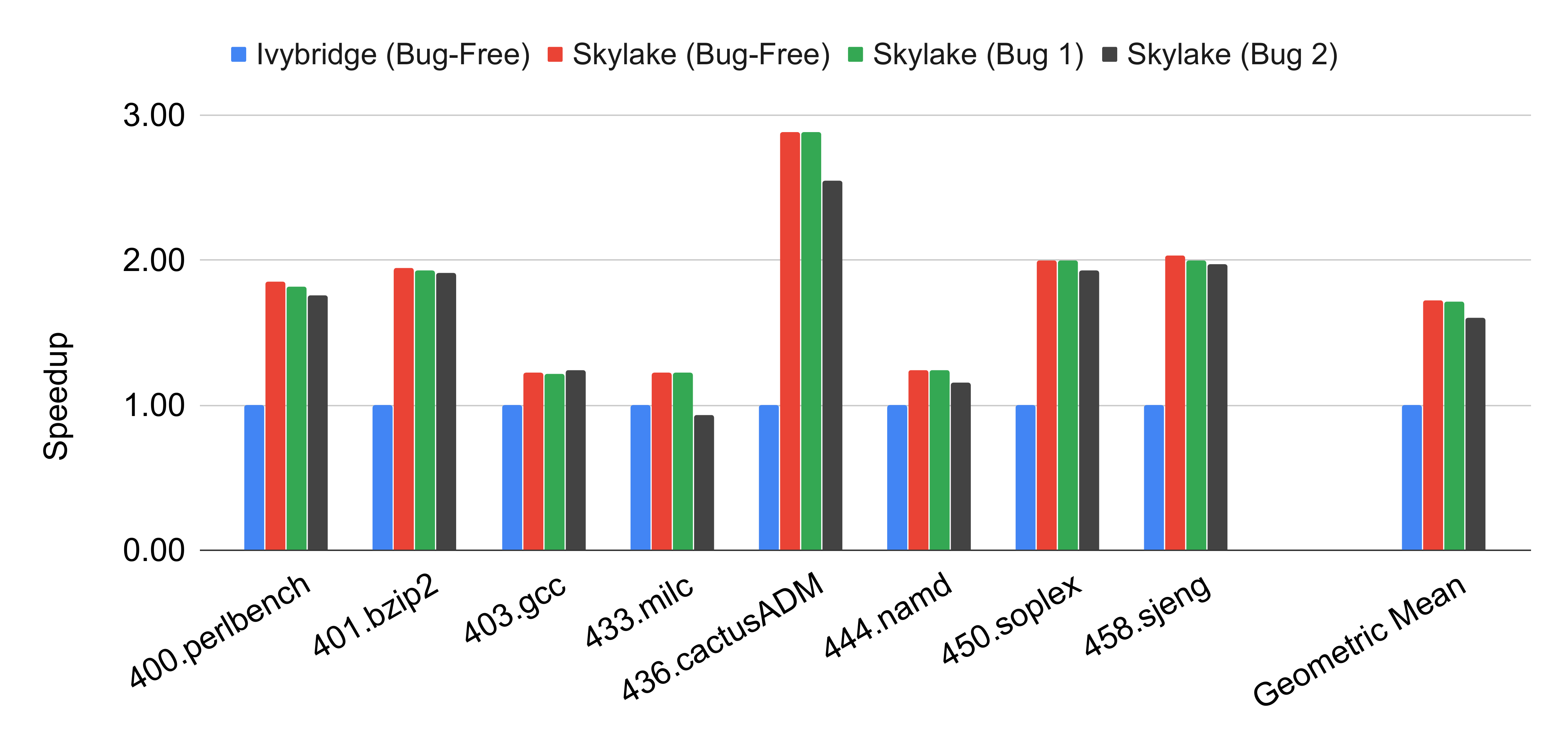}
     \caption{Speedup of Skylake simulation with and without performance bugs,
       normalized against Ivybridge simulation.}
    \label{fig:ivy_vs_sky}
\end{figure}

Performance bugs, while potentially significant versus a bug-free
version of the given microarchitecture, may be less than the
difference between microarchitectures.  Consider the data shown in
Figure~\ref{fig:ivy_vs_sky}.  The figure shows the performance of
several SPEC CPU2006 benchmarks~\cite{spec2006} on gem5~\cite{gem5},
configured to emulate two recent Intel microarchitectures: Ivybridge
(3$^{rd}$ gen Intel ``Core''~\cite{ivybridge}) and Skylake (6$^{th}$
gen~\cite{skylake}), using the methodology discussed in
Sections~\ref{sec:method} and~\ref{sec:experiment}.
The figure shows that the baseline, bug-free Skylake microarchitecture
provides nearly 1.7X the performance of the older Ivybridge, as
expected given the greater core execution resources provided.

In addition to a bug-free case, we collected performance data for two
arbitrarily selected bug-cases for Skylake. \emph{Bug 1} is an
instruction scheduling bug wherein \textit{xor} instructions are the
only instructions selected to be issued when they are the oldest
instruction in the instruction queue.
As the figure shows, the
overall impact of this bug is very low ($<1\%$ in average
across the given workloads). \emph{Bug 2} is another instruction
scheduling bug which causes \textit{sub} instructions to be
incorrectly marked as synchronizing, thus all younger instructions
must wait to be issued after any \textit{sub} instruction has been
issued and all older instructions must issue before any \textit{sub}
instruction, regardless of dependencies and hardware resources.  The
impact of this bug is larger, at $\sim7\%$ across the given workloads.

Nevertheless, for \emph{both} bugs, the impact is less than the
difference between Skylake and Ivybridge.  Thus, from the designers
perspective, working on the newer Skylake design with Ivybridge in
hand, if early performance models showed the results shown for
\emph{Bug 2} (even more so \emph{Bug 1}) designers might incorrectly
assume that the buggy Skylake design did not have performance bugs
since it was much better than previous generations. Even as the
performance difference between microarchitecture generations
decreases, differentiating between performance bugs and expected
behavior remains a challenge given the variability between simulated
performance and real systems.

Examining the figure, we note that each bug's impact varies 
from workload to workload.  That said, in the case of \emph{Bug 1}, no
single benchmark shows a performance loss of more than 1\%.  This
makes sense since the bug only impacts the performance of
applications which have, relatively rare, \emph{xor} instructions.
From this, one might conclude that \emph{Bug 1} is trivial and perhaps
needs not be fixed at all.  This assumption however, would be wrong, as
many workloads do in fact rely heavily upon the \emph{xor}
instruction, enough so that a moderate impact on them translates into
significant performance impact overall.

In this work, we explore several approaches to this problem
and ultimately develop an automated, two-stage, machine learning-based
methodology that will allow designers to automatically detect the
existence of performance bugs in microprocessor designs. Even though this
work focus is the processor core, we show that the methodology 
can be used for other system modules, such as the memory. 
\shepherd{Our approach extracts and exploits knowledge from legacy designs instead
of relying on a reference model. Legacy designs have already undergone major debugging,
therefore we considered them to be bug-free, or to have only minor performance bugs.}
The automatic detection 
takes several minutes of machine learning inference
time and several hours of architecture simulation time.  The major
contributions of this work include the following:
\begin{compactitem}
\itemsep0em 
\item This is the first study on using machine learning for
  performance bug detection, to the best of our knowledge. \hl{Thus,
    while we provide a solution to this particular problem, we also
    hope to draw the attention of the research community to the broader performance validation domain.}
\item Investigation of different strategies and 
  machine learning techniques to approach the problem.  Ultimately, an
  accurate and straight forward, two phase approach is developed.
\item \begin{sloppypar} A novel automated method to leverage
  SimPoints~\cite{calder2005simpoint} to extract, short,
  per\-for\-mance-\-orthogonal, microbenchmark ``probes'' from long
  running workloads.\end{sloppypar}
\item A set of processor performance bugs which may be configured for
  an arbitrary level of impact, for use in testing
  performance debugging techniques is developed.
\item \hlcyan{Our methodology detects 91.5\% of these processor core performance bugs 
  leading to $\geq 1\%$ IPC degradation with 0\% false positives.
  It also achieves 100\% accuracy for the evaluated performance bugs in
  memory systems.}
\end{compactitem}

As an early work in this space, we limit our scope  somewhat
  to maintain tractablity. \hlcyan{This work is focused mainly 
  on the processor
  core (the most complex component in the system), however we show
  that the methodology can be generalized to other units by evaluating
  its usage on memory systems.}
  Further, here we focus on
  performance bugs that affect cycle count and consider timing issues to
  be out of scope.  
  
  \shepherd{The goal of this work is to find bugs in new architectures that
  are incrementally evolved from those used in the training data, it may not work
  as well when there have been large shifts in microarchitectural details, such as
  going from in-order to out-of-order.} However, major generalized 
  microarchitectural changes are increasingly rare. For example, Intel processors’
  last major microarchitecture refactoring occurred $\sim15$ years ago, from
  P4 to  ``Core''. Over this period, there have been several microarchitectures 
  where our approach could provide value. In the event of larger changes, our method can be partially reused by augmenting it with additional features.




\section{Objective and A Baseline Approach}
\label{sec:baseline}

The \emph{objective} of this work is the detection of
performance bugs in new microprocessor 
designs. We 
explicitly look at microarchitectural-level bugs, as opposed
to logic or architectural-level bugs.  As grounds for the
detection, one or multiple legacy microarchitecture designs are
available and assumed to be bug-free or have only minor
  performance bugs remaining.  Given the extensive pre-silicon and
  post-silicon debug, this assumption generally
  holds on legacy designs.  There are no assumptions
  about how a bug may look like.  As it is very difficult to define a theoretical coverage model for 
  performance
  debug, our methodology is a heuristic approach.
  
We introduce a na\"ive baseline approach to solving the bug detection
problem.  This serves as a prelude for describing our proposed
methodology. It also provides a reference for comparison in
the experiments.

Performance bug detection bears certain similarity to standard
chip functional bug detection. Thus, a testbench strategy similar
to functional verification is sensible. That is, a set of
microbenchmarks are simulated on the new microarchitecture, and
certain performance characteristics are monitored and analyzed.  The
key difference is that functional verification has correct responses
to compare against while bug-free performance for a new
microarchitecture is not well defined.

The baseline approach uses supervised machine learning to
classify if an microarchitecture has performance bugs or not.  The
input features for a machine learning model consist of performance
counter data, IPC (committed Instructions Per Cycle) and
microarchitecture design parameters, such as cache sizes.  One
classification result on bug existence or not is produced for each
application.  The overall detection result is obtained by a
voting-based ensemble of classification results among multiple
applications.  Let $\rho$ be the ratio of the number of applications
with positive classification (indicating bug) versus the total number
of applications. A microarchitecture design is judged to have a bug
if $\rho \ge \theta$, where $\theta$ is a threshold determining the
tradeoff between true positive  and false positive rates. The
models are trained by legacy microarchitecture designs and bugs are
inserted to obtain labeled data.

The details of this approach, such as applications,
machine learning engines, performance counter selection,
microarchitecture parameters and bug development,
overlap with our proposed method and will be described in later
sections.

\section{Proposed Methodology}
\label{sec:method}

\subsection{Key Idea and Methodology Overview}
\label{sec:overview}

\begin{figure}[htb!]
  \centering
  \includegraphics[width = 0.4\textwidth]{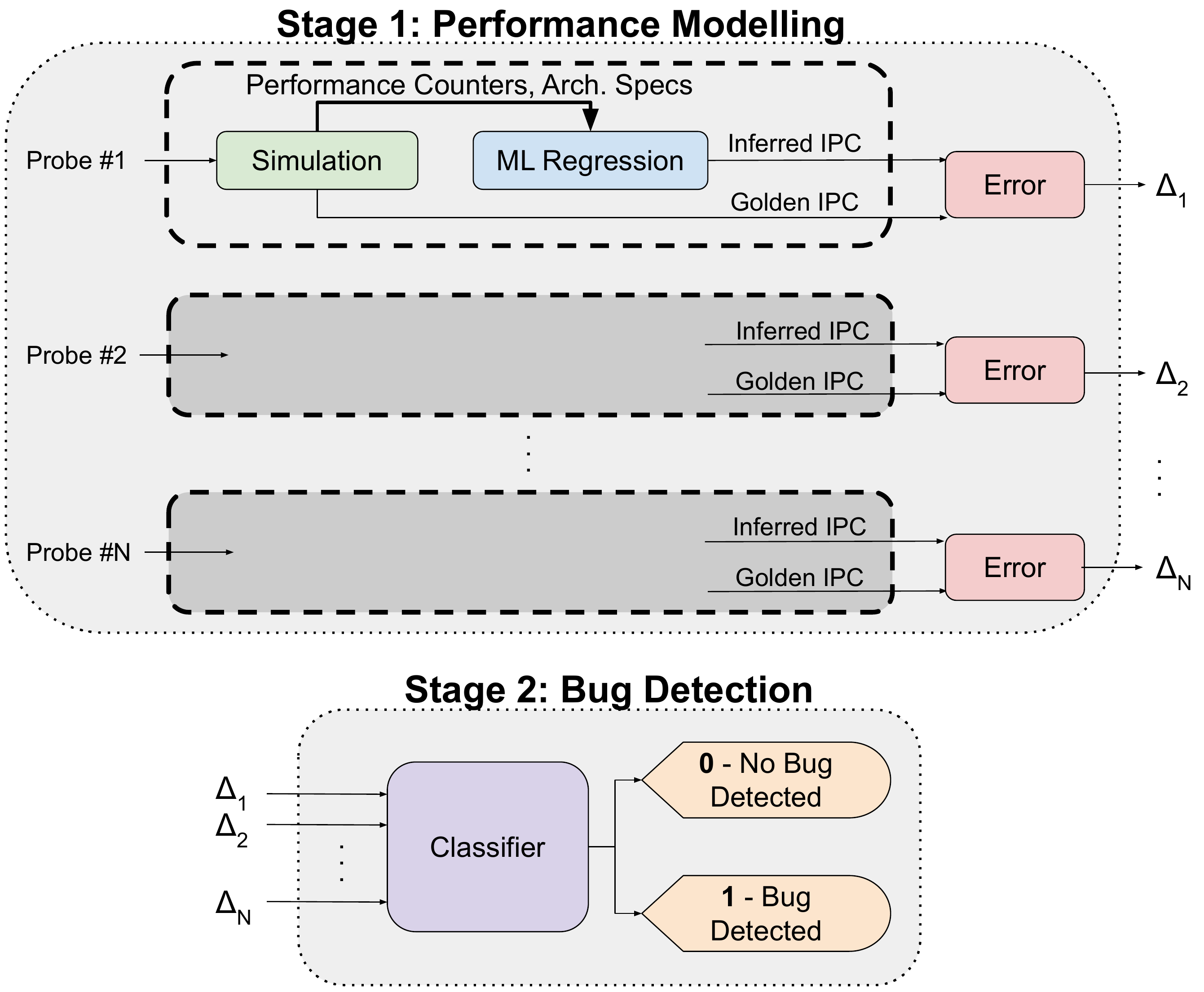}
  \caption{Overview of the proposed methodology.}
  \label{fig:flow_diagram_method}
\end{figure}

The proposed methodology is composed of two stages.  The first stage
consists of a machine learning model trained with performance counter
data from bug-free microarchitectures to infer microprocessor
performance in terms of IPC.  If such model is well-trained and
achieves high accuracy, it will capture the correlation between the
performance counters and IPC in bug-free microarchitectures.  When
this model is then applied on designs with bugs, a significant
increase in inference errors is expected, as the bugs ruin the
correlation between the counters and IPC. The second stage
determines bug existence due to the IPC inference errors in the first
stage.  An overview of the two-stage methodology is depicted in
Figure~\ref{fig:flow_diagram_method}. 
\shepherd{Although in this work we evaluate the methodology using only
microarchitecture simulations, its key idea is generic and can be extended
for FPGA prototyping or post-silicon testing. 
The applicability in both pre- and post-silicon validations is 
a strong point of our methodology, especially as some bugs might be
triggered by traces that are too long to be simulated.
}


In this methodology, a micro\-bench\-mark and a set of selected
performance counters form a \textbf{performance probe}.  Probe design
is introduced in Section~\ref{subsec:probes}.  The machine learning
model is applied on individual probes for IPC inference, which is
described in Section~\ref{subsec:method_stg_1}.  The classification in
stage two is based on the errors from multiple probes, and elaborated
in Section~\ref{sub:bug_detection}.

\subsection{Performance Probe Design} \label{subsec:probes}

\subsubsection{SimPoint-Based Microbenchmark Extraction}
\label{sec:simpoint}

\begin{figure*}[!htb]
    \centering
     \includegraphics[width = 0.66\textwidth]{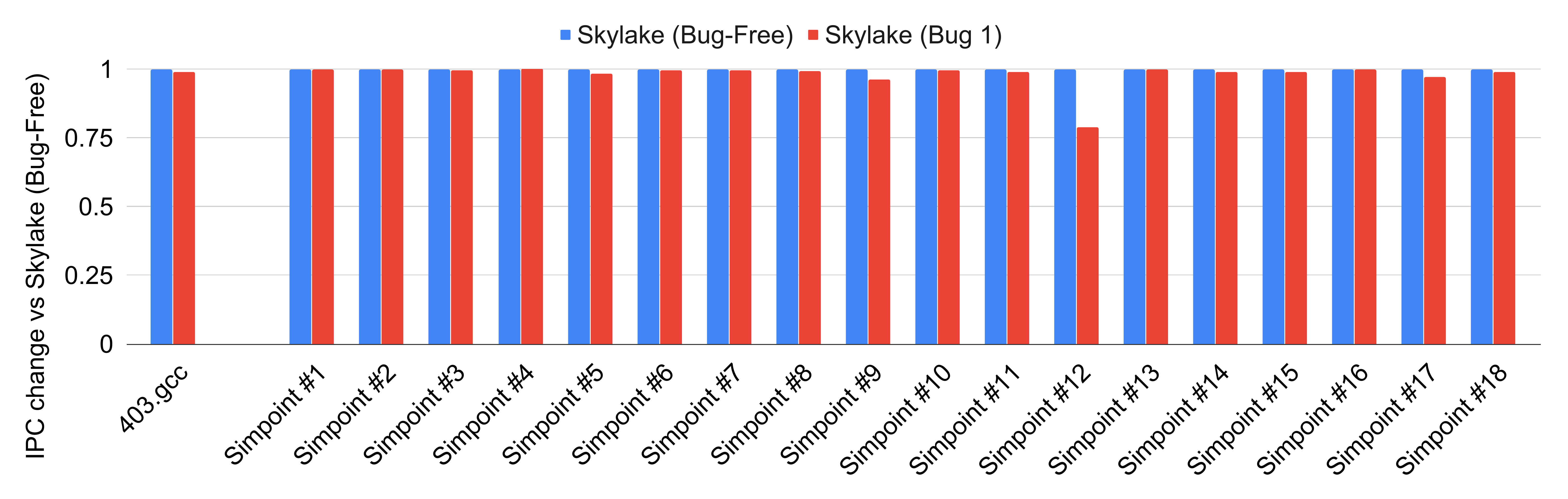}
    \caption{IPC by SimPoints in 403.gcc for Skylake architecture.}
    \label{fig:sky_gcc}
\end{figure*}

Finding the right applications to use in ``probing'' the
microarchitecture for performance bugs is critical to the efficiency
and efficacy of the proposed methodology.  One possible approach could
be to meticulously hand generate an exhaustive set of microbenchmarks
to test all the interacting components of the microarchitecture under
all possible instruction orderings, branch paths, dependencies,
\emph{etc}.  This approach is similar to that taken in some prior
works~\cite{delimitrou2013ibench,leverich2014reconciling,delimitrou2015tarcil}.
While this approach would likely be possible, automating the process
as much as possible is highly desirable, given the overheads already
required for verification and validation. Thus, another key goal of
our work is to automatically find and select short, orthogonal and relevant
performance micro\-benchmarks to use in the
microarchitecture probing.

Typically in computer architecture research, large, orthogonal suites
of workloads, such as the SPEC CPU suites~\cite{spec2006,spec2017},
are used for performance bench\-marking.  The applications in these
suites, however, typically are too large to simulate to completion.
Thus, development of statistically accurate means to shorten
  benchmark runtimes has received significant
  attention~\cite{Conte1996, wunderlich2003smarts,
  perelman2003using}.  One notable
work 
was
SimPoints~\cite{perelman2003using,hamerly2004use,calder2005simpoint},
wherein the authors propose to identify short, performance-orthogonal
segments of long running applications.  These segments 
are simulated separately and the whole application performance is
estimated as a weighted average of the performance of those points.
In this work, we propose a novel use of the SimPoints methodology.
Instead of using them as intended, for performance estimation of
large applications, we propose to use
them 
to \emph{automatically} identify and extract short, orthogonal,
performance-relevant microbenchmarks from long running applications.
Here we are leveraging SimPoints' identification of orthogonal
basic-block vectors in the given program to be the source of our
orthogonal microbenchmarks.


As an example how this provides greater visibility into performance
bugs, consider Figure \ref{fig:sky_gcc}.  In the figure, we compare
the performance per SimPoint of bug-free and \emph{Bug 1} versions
of Skylake (described in Section~\ref{sec:intro}) for the benchmark
403.gcc.  We see that, although the overall difference in the whole
application is less than 1\%, when we split the performance by
SimPoint, SimPoint \#12 has a
degradation of over 20\%, making \emph{Bug 1}'s behavior much easier
to identify as incorrect.

The reason why SimPoint \#12 shows this behavior is
because 
it 
has more than twice the number of \textit{xor} operations than the
others, accounting for 2.3\% of all instructions executed
here.  
We argue
that, even though the overall impact on the performance in 403.gcc is
very low, this particular SimPoint represents well a class of
applications with similar behavior that would not be represented by
looking at any single benchmark in the SPEC CPU suite. Thus, by
utilizing SimPoints individually more performance bug coverage can be
gained than by looking at full application performance.

\subsubsection{Performance Counter Selection} \label{subsec:counters}

Performance counter data from micro\-bench\-mark simulation constitute
the main features for our machine learning models to infer overall
performance. 
Hence, they are essential components for the performance probes.
We note that, as part of the design cycle, performance counters
  often undergo explicit validation. 
  The assumption of their sanity is central for performance debug in
  general, not only this methodology. 

Microprocessors typically contain hundreds to thousands of performance
counters.  Using all of them as features would unnecessarily extend
training and inference time for machine learning models, and even
degrade them as some useless counters may blur the picture.  Hence, we
select a small subset of them for each microbenchmark that is
sufficient for effective machine learning inference on that
workload. We note that the abundance of counters along with our
  selection method makes our methodology resilient to small changes in
  the counters available across different generations of architecture
  designs.

The selection is carried out in two steps.
\begin{compactitem}
\item Step 1: The Pearson correlation coefficient is evaluated for
  each counter with respect to IPC.  Only those counters with high
  correlations with IPC (magnitude greater than 0.7) are retained and
  the others are pruned out.
\item Step 2: Among the remaining counters, the correlation between
  every pair of them is evaluated.  If two counters have a Pearson
  correlation coefficient whose magnitude is greater than 0.95, we
  consider them redundant and one of them is pruned out.
\end{compactitem}

Counter selection is conducted independently for each probe
and thus different probes may use different counters.  Examples of the
most commonly selected counters are: the number of fetched
instructions, percentage of branch instructions, number of writes to
registers, percentage of correctly predicted indirect branches, number
of cycles on which the maximum possible number of instructions were
committed, among others.

\subsection{Stage 1: Performance Modelling} \label{subsec:method_stg_1}

Stage 1 of our methodology models overall processor performance
(IPC), via machine learning. Among innumerable
possible ways to perform such task, we elaborately develop one with the following key ingredients:
\begin{compactenum}
\itemsep0em 

\item The training data are taken from (presumably) bug-free designs.
Otherwise, the inference errors between
  bug-free designs and bug cases cannot be differentiated.
\item One model is trained for each probe.  As workload
  char\-ac\-ter\-is\-tics of different probes can be drastically
  different, it is very difficult (also unnecessary), for a single
  model to capture them all.  A separated model for each probe is
  necessary to achieve high accuracy for bug-free designs. This is a
    key difference from previous methods of performance counter-based
    predictions
where a single model is derived to work across many different workload~\cite{Joseph2001,Contreras2005,Bircher2007,Yang2016}.
By tuning our model to the particular
    counters relevant to a particular probe workload, much higher
    accuracy can be achieved.
\item Models are trained with a variety of
  microarchitectures.  This ensures that inference error
  differences are due to existence of bugs instead of
  microarchitecture difference.  In practice, the training is
  performed on legacy designs and the model inference would be on a
  new microarchitecture.
\item The input features are taken as a time series. The simulation
  (or running system) proceeds over a series of time steps
  $t_1, t_2, ...$.  Let the current time step be $t_i$, then the model
  takes feature data of $t_{i-w+1}, t_{i-w+2}, ..., t_{i-w+w}$ as
  input to estimate IPC at $t_i$, where $w$ is the window size.
  Usually, a time step consists of multiple clock cycles, e.g., a half
  million cycles.  If one intends to take the aggregated data of an
  entire SimPoint as input, that is a special case of our framework,
  where the step size is the entire SimPoint and the window size is 1.
\end{compactenum}

Besides the selected counters discussed in
Section~\ref{subsec:counters}, some microarchitecture design parameters are
used as features to our machine learning model.  They include clock
cycle, pipeline width, re-order buffer size and some cache
characteristics such as latency, size and associativity.  Please note
that the parameter features are constant in the time series input for
a specific microarchitecture. 

The following machine learning engines are investigated for per-probe
IPC inference in this work
\vskip 0.45pt

\emph{Linear Regression (\textbf{Lasso})}
This is a simple linear regression algorithm
\cite{santosa1986linear}, of the form
  $y = \mathbf{x}^T \mathbf{w}$, where $\mathbf{w}$ is the vector of
  parameters.  It uses L1 regularization on the values of
  $\mathbf{w}$. The main advantage of this model is its simplicity.
\vskip 0.45pt

\emph{Multi-Layer Perceptron (\textbf{MLP})} This classical neural
network consists of multiple layers of perceptrons.  Each perceptron
takes a linear combination of signals of the previous layer and
generates an output via a non-linear activation
function~\cite{hornik1989multilayer}.
\vskip 0.45pt

\emph{Convolutional Neural Network (\textbf{CNN})}
This neural network architecture contains convolution
operations and the parameters of its convolution kernel are trained with
data~\cite{lecun1995convolutional}.  In this work, the features are
taken as 1D vector as described by Eren \emph{et
  al.}~\cite{eren2019generic} and Lee \emph{et
  al.}~\cite{lee2017human}, as opposed to 2D images.
\vskip 0.45pt

\emph{Long Short-Term Memory Network (\textbf{LSTM})}
This is a form of recurrent neural network, where the output depends
not only on the input but also internal
states~\cite{hochreiter1997lstm}.  The states are useful in capturing
history effect and make LSTM effective for handling time series
signals.
\vskip 0.45pt

\emph{Gradient Boosted Trees (\textbf{GBT})}
This is a decision tree-based approach, where a branch node in the
tree partitions the feature space and the leaf nodes tell the
inference results. The overall result is an ensemble of results from
multiple trees~\cite{friedman2001greedy}. In this work, we adopt
XGBoost~\cite{chen2016xgboost}, which is the latest generation of GBT
approach.

Consider a set of probes $P=\{p_1, p_2, ... \}$.  A trained model is
applied to infer IPC $\hat{y}_{i,j}$ for time step $t_j$ of probe
$p_i$ for a particular design, while the corresponding IPC by
simulation is $y_{i,j}$. For each probe $p_i \in P$, the inference
error is 
\begin{equation}
    \Delta_i = \frac{1}{2} \sum_{j=2}^{T_i} \left(|y_{i,j} - \hat{y}_{i,j}| + |y_{i,j-1} - \hat{y}_{i,j-1}| \right) \label{eq:error}
\end{equation}
where $T_i$ is the number of time steps in $p_i$. This error can be
interpreted as the area of difference between the inferred IPC over
time and the actual (simulated) IPC.  It is also approximately equal
to the total absolute error.  The set of errors
$\{\Delta_1, \Delta_2, ..., \Delta_{|P|}\}$ are fed to stage 2 of our
methodology.  Empirically, this error metric outperforms others, such
as Mean Squared Error (MSE), in the bug classification of stage 2.
An advantage of error metric \eqref{eq:error} is that a large
  error in a single (or few) time step(s) is not averaged out, as it
  is with MSE.

  \shepherd{Our machine learning models are not intended to be golden
    performance models. Instead, each model captures the complex
    relationship between counters and performance for the workload in
    exactly one probe. Their comparison against the simulated
    performance for that particular probe is not to determine if
    performance is satisfactory, rather it is to attempt to detect if
    the model accuracy is ruined, as this might signal the relations
    between counters and performance are broken, therefore a
    performance bug is likely. The usage of machine learning is
    essential for bug detection as it overcomes the difficulty to
    obtain a single, accurate, universal bug-free performance model.
  }

\subsection{Stage 2: Bug Detection} \label{sub:bug_detection}

Given a vector $[\Delta_1 , \Delta_2 , \cdots , \Delta_{|P|}]$ of IPC
inference errors from stage 1, the purpose of stage 2 is to identify
if a bug exists in the corresponding microarchitecture.  In general, a
small $\Delta_i$ error indicates that the machine learning model in
stage 1, which is trained with bug-free designs, matches well the
design under test and hence likely no bugs (with respect to that
probe).  On the other hand, large $\Delta_i$ error indicates a large
likelihood of bugs.  Bugs can reside at different locations and
manifest in a variety of ways.  Therefore, multiple orthogonal probes
are necessary to improve the breadth and robustness of the detection,
as highlighted in the case of \emph{Bug 1} in
Figure~\ref{fig:sky_gcc}.

\shepherd{Although we considered using a machine learning classifier
  for stage 2, we found the available training data for this stage to
  be much more scarce than that of stage 1.  For stage 1, every time
  step of the collected data 
  represents a data sample, making thousands of samples available. In
  contrast, only one data sample per microarchitecture is available
  for stage 2, making only a few dozens of samples available. To
  overcome the lack of data available for training we developed a
  custom rule-based classifier.  }

Suppose there are $m$ positive samples (with bugs) and $n$ negative
samples (bug-free). For each probe $p_i \in P$, we compute the mean
$\mu_i^{+}$ and standard deviation $\sigma_i^{+}$ of the IPC
inference errors among the $m$ positive samples.  Similarly, we can
obtain $\mu_i^{-}$ and $\sigma_i^{-}$ among the $n$ negative samples
for each $p_i \in P$.  These statistics form a reference to evaluate
the IPC inference errors for a new architecture.



For the error $\Delta^\prime_i$ of applying probe $p_i$ on the new
architecture, we evaluate the following ratios based on the statistics
of the labeled data.
\begin{equation}
  \label{eq:gammas}
  \gamma_i^+ = \frac{\Delta^\prime_i}{\mu_i^+ + \alpha \sigma_i^+} ~ , ~~
  \gamma_i^- = \frac{\Delta^\prime_i}{\mu_i^- + \alpha \sigma_i^-}
\end{equation}
where $\alpha$ is a parameter trained with the labeled data. In general,
a large value of $\Delta^\prime_i$ signals a high probability of bugs in
the new microarchitecture.  Ratios $\gamma_i^+$ and $\gamma_i^-$ are
relative to the labeled data, and make the errors from different
probes comparable.

Given vectors $[\gamma_1^+, \gamma_2^+, ..., \gamma_{|P|}^+]$ and
$[\gamma_1^-, \gamma_2^-, ..., \gamma_{|P|}^-]$, our classifier is a
rule-based recipe as follows.

\begin{compactenum}
\item If $\max(\gamma_1^+, \gamma_2^+, ..., \gamma_{|P|}^+) > \eta$,
  where $\eta$ is a parameter, this architecture is classified to have
  bugs. This is for the case where a large error appears in at least
  one probe regardless of errors from the other probes.
\item If
  $(\gamma_1^-+ \gamma_2^-+ ...+ \gamma_{|P|}^-)/|P| > \lambda$, where
  $\lambda < \eta$ is a parameter, this architecture is classified to
  have bugs. This is for the case where small errors appear on many
  probes.
\item For other cases, the architecture is classified as bug-free.
\end{compactenum}



While the values of $\eta$ and $\lambda$ are empirically chosen as 15
and 5, respectively, the value of $\alpha$ is trained according to
true positive rate and false positive rate on the labeled data. In the
training, a set of uniformly spaced values in a range are tested for
$\alpha$, and the one with the maximum true positive rate and false
positive rate no greater than 0.25 is chosen.


\section{Experimental Setup}
\label{sec:experiment}

\hlcyan{In this section we elaborate on the details of the implementation
of our methodology. Sections }\ref{subsec:probe_setup} to \ref{subsec:bug_setup}
\hlcyan{cover the experimental setup for processor core performance bugs, since this 
is the main focus of our work, we provide detailed explanations. 
Section } \ref{subsec:detection_mem} 
\hlcyan{lists the changes implemented for the memory system performance bug detection.}

\subsection{Probe Setup}
\label{subsec:probe_setup}

The performance probes are extracted from 10 applications from the
SPEC CPU2006 benchmark~\cite{spec2006} using the SimPoints
methodology~\cite{calder2005simpoint}, as described in
Section~\ref{sec:simpoint}. We chose to use SPEC CPU2006 instead of
the more recent SPEC CPU2017 suite because their relatively smaller
memory footprints, generally reduced running times and computational
resources needed to detect performance bugs, though there is no reason
our techniques would not work with SPEC CPU2017 applications.  Unlike
developing performance improvement techniques, where the benchmarks
serve as testcases and should be exhaustively used, the selected
applications here are part of our methodology implementation.
In fact, the selection affects the tradeoff between the efficacy of
detection and runtime overhead of the methodology, as we will show.

There are 190 SimPoints in total for the ten SPEC CPU2006 applications
we use here and each SimPoint contains around 10M instructions. The
applications we selected in particular were an arbitrary set of the
first 10 we were able to compile and run in gem5 across all
microarchitectures.  Adding more benchmarks would only improve the
results we achieve.  A list of the applications is provided in
Table~\ref{tab:probes}.

For each probe, between 4 and 64 performance counters are selected
using the methodology detailed in Section~\ref{subsec:counters}. The
time step size is 500k clock cycles. That is, for each counter,
a value is recorded every 500k clock cycles. The values of all
time steps form the time series as input feature to the machine
learning models. As the time step size is sufficiently large, the
default window size (see Section~\ref{subsec:method_stg_1}) is one
time step.

\begin{scriptsize}
\begin{table}[!htb]
\centering
\caption{Selected SPEC CPU2006 benchmarks.}
\resizebox{0.47\textwidth}{!}{
\begin{tabular}{c|c|c|c}
\hline
\textbf{Benchmark} & \textbf{Operand Type} & \textbf{Application} & \textbf{SimPoints} \\ \hline
400.perlbench & Integer & PERL Programming Language & 14 \\
401.bzip2 & Integer & Compression & 23 \\
403.gcc & Integer & C Compiler & 18 \\
426.mcf & Integer & Combinatorial Optimization & 15 \\
433.milc & Floating Point & Quantum Chromodynamics & 20 \\
436.cactusADM & Floating Point & General Relativity & 16 \\
444.namd & Floating Point & Molecular Dynamics & 26 \\
450.soplex & Floating Point & Linear Programming & 21 \\
458.sjeng & Integer & AI: Chess & 19 \\
462.libquantum & Integer & Quantum Computing & 18\\
\hline
\end{tabular}
}
\label{tab:probes}
\end{table}
\end{scriptsize}


\subsection{Simulated Architectures} \label{subsec:sim_archs}

All experiments are performed via gem5~\cite{gem5} simulations in
System Emulation mode. Note, the key ideas of our methodology can
  be implemented in other simulators and post-silicon debug.  Here we
are focused on core performance bugs, thus we use the Out-Of-Order
core (O3CPU) model with the x86 ISA.  
\shepherd{Gem5 is configured to model different architectures by varying
several configuration variables such as clock period, re-order buffer size, 
cache size, associativity, latency, and number of levels, as well as functional
unit count, latency, and port organization. 
Besides these common knobs, other microarchitectural differences between designs
are not considered.
The configurations model eight different existing microarchitectures: Intel Broadwell,
Cedarview\footnote{Implements a Cedarview-like superscalar architecture but assumes out-of-order completion}, Ivybridge, Skylake and Silvermont, AMD
Jaguar, K8, and K10, as well as 12 artificial microarchitectures
with realistic settings. With these, we achieve a wide variety of architectures,
from low-power systems such as Silvermont to high-performance server systems such as Skylake.}
Detailed information of the specifications used for each architecture can be found
in Table \ref{tab:arch_specs}. Table \ref{tab:port_archs} describes the port organization
for each of the architectures.
These microarchitectures serve as
training data for machine learning models and unseen data for testing
the models.
They are partitioned into
four disjoint sets as follows.
\begin{compactitem}
\itemsep0em 
\item \textbf{Set I:} Contains two real and
  seven artificial microarchitectures. They are used to train our IPC
  models.
\item \textbf{Set II:} Three microarchitectures in this set, a real
  one and two artificial ones, play the role of validation in training
  the IPC models.  That is, a training process terminates
  after 100 epochs of training without improvement on this set. In
  addition, this set serves as labeled data for training the classifier in
  stage 2.
\item \textbf{Set III:} Four microarchitectures, one real and three
  artificial, are used as additional training data for the stage 2
  classifier. As such, the training data for stage 2 is composed by set
  II and III. This is to decouple the training data between stage 1
  and stage 2.
\item \textbf{Set IV:} Four microarchitectures in this set are for the
  testing of the stage 2 classifier. As stage 2 is the eventual bug
  detection, all of the four microarchitectures here are real ones so
  as to ensure the overall testing is realistic.
\end{compactitem}

\begin{scriptsize}
\begin{table*}[!htb]
\centering
\caption{Architectural knobs for implemented architectures.}
\label{tab:arch_specs}
\resizebox{\textwidth}{!}{
\begin{tabular}{ccccccccc}
\hline
\textbf{Set} & \textbf{Architecture} & \multicolumn{1}{c}{\textbf{\begin{tabular}[c]{@{}c@{}}Clock\\ Cycle\end{tabular}}} & \multicolumn{1}{c}{\textbf{\begin{tabular}[c]{@{}c@{}}CPU\\ Width\end{tabular}}} & \multicolumn{1}{c}{\textbf{\begin{tabular}[c]{@{}c@{}}ROB\\ Size\end{tabular}}} & \multicolumn{1}{c}{\textbf{\begin{tabular}[c]{@{}c@{}}L1 Cache\\ (Size / Assoc. / Latency)\end{tabular}}} & \multicolumn{1}{c}{\textbf{\begin{tabular}[c]{@{}c@{}}L2 Cache\\ (Size / Assoc. / Latency)\end{tabular}}} & \multicolumn{1}{c}{\textbf{\begin{tabular}[c]{@{}c@{}}L3 Cache\\ (Size / Assoc. / Latency)\end{tabular}}} & \multicolumn{1}{c}{\textbf{\begin{tabular}[c]{@{}c@{}}Func. Unit Latency\\ (FP / Multiplier / Divider)\end{tabular}}} \\
\hline
I & Broadwell & 4.0GHz & 4 & 192 & 32kB / 8-way / 4 cycles & 256kB / 8-way / 12 cycles & 64MB / 16-way / 59 cycles & 5 cycles / 3 cycles / 20 cycles \\
I & Cedarview & 1.8GHz & 2 & 32 & 32kB / 8-way / 3 cycles & 512kB / 8-way / 15 cycles & No L3 & 5 cycles/ 4 cycles / 30 cycles \\
I & Jaguar & 1.8GHz  & 2 & 56 & 32kB / 8-way / 3 cycles & 2MB / 16-way / 26 cycles & No L3 & 4 cycles / 3 cycles / 20 cycles \\
I & Artificial 2 & 4.0GHz & 8 & 168 & 32kB / 2-way / 5 cycles & 256kB / 8-way / 16 cycles & No L3 & 4 cycles / 4 cycles / 20 cycles \\
I & Artificial 3 & 3.0GHz & 8 & 32 & 32kB / 2-way / 3 cycles & 512kB / 16-way / 24 cycles & 8MB / 32-way / 52 cycles &  4 cycles / 4 cycles / 20 cycles \\
I & Artificial 4 & 4.0GHz & 2 & 192 & 64kB / 8-way / 3 cycles & 1MB / 8-way / 20 cycles & 32MB / 16-way / 28 cycles & 5 cycles / 3 cycles / 20 cycles \\
I & Artificial 6 & 3.5GHz & 4 & 192 & 64kB / 8-way / 4 cycles & 1MB / 8-way / 16 cycles & 8MB / 32-way / 36 cycles & 4 cycles / 4 cycles / 20 cycles \\
I & Artificial 7 & 3.0GHz & 4 & 32 & 16kB / 8-way / 3 cycles & 512kB / 16-way / 12 cycles & 32MB / 32-way / 28 cycles & 2 cycles / 7 cycles / 69 cycles \\
I & Artificial 10 & 1.5GHz & 8 & 32 & 32kB / 2-way / 2 cycles & 256kB / 16-way / 24 cycles & 64MB / 32-way / 36 cycles & 5 cycles/ 4 cycles / 30 cycles \\
I & Artificial 11 & 3.5GHz & 4 & 32 & 64kB / 4-way / 5 cycles & 256kB / 4-way / 24 cycles & No L3 & 5 cycles/ 4 cycles / 30 cycles \\

II & Ivybridge & 3.4GHz & 4 & 168 & 32kB / 8-way / 4 cycles & 256kB / 8-way / 11 cycles & 16MB / 16-way / 28 cycles & 5 cycles / 3 cycles / 20 cycles  \\
II & Artificial 0 & 2.5GHz & 4 & 192 & 64kB / 2-way / 4 cycles & 512kB / 4-way / 12 cycles & No L3 & 5 cycles / 3 cycles / 20 cycles \\
II & Artificial 9 & 3.5GHz & 8 & 192 & 16kB / 4-way / 5 cycles & 1MB / 4-way / 20 cycles & 64MB / 16-way / 44 cycles & 4 cycles / 3 cycles / 11 cycles \\

III & Artificial 1 & 1.5GHz & 4 & 192 & 64kB / 8-way / 5 cycles & 2MB / 8-way / 16 cycles & No L3 & 4 cycles / 3 cycles / 11 cycles \\
III & Artificial 5 & 3.5GHz & 2 & 32 & 32kB / 4-way / 5 cycles & 256kB / 4-way / 16 cycles & 8MB / 32-way / 44 cycles & 4 cycles / 3 cycles / 11 cycles \\
III & Artificial 8 & 3.0GHz & 2 & 192 & 32kB / 2-way / 2 cycles & 1MB / 16-way / 16 cycles & 32MB / 32-way / 52 cycles & 4 cycles / 3 cycles / 11 cycles \\

IV & K8 & 2.0GHz & 3 & 24 & 64kB / 2-way / 4 cycles & 512kB / 16-way / 12 cycles & No L3 &  4 cycles / 3 cycles / 11 cycles \\
IV & K10 & 2.8GHz & 3 & 24 & 64kB / 2-way / 4 cycles & 512kB / 16-way / 12 cycles & 6MB / 16-way / 40 cycles & 4 cycles / 3 cycles / 11 cycles \\
IV & Silvermont & 2.2GHz & 2 & 32 & 32kB / 8-way / 3 cycles & 1MB / 16-way / 14 cycles & No L3 & 2 cycles / 7 cycles / 69 cycles \\
IV & Skylake & 4.0GHz & 4 & 256 & 32kB / 8-way / 4 cycles & 256kB / 4-way / 12 cycles & 8MB / 16-way / 34 cycles & 4 cycles / 4 cycles / 20 cycles \\

\hline
\end{tabular}
}
\end{table*}
\end{scriptsize}

\begin{scriptsize}

\begin{table*}[!htb]
\centering
\caption{Port organization of implemented architectures.}
\label{tab:port_archs}
\resizebox{\textwidth}{!}{%
\begin{tabular}{ccccccccc}
\hline
\textbf{Set} & \textbf{Architecture} & \textbf{Port 0} & \textbf{Port 1} & \textbf{Port 2} & \textbf{Port 3} & \textbf{Port 4} & \textbf{Port 5} & \textbf{Port 6} \\ \hline
I & Broadwell & \begin{tabular}[c]{@{}c@{}}1 ALU, 1 FP Mult\\ 1 FP unit, 1 Int Vector\\ 1 Int Mult, 1 Divider\\ 1 Branch Unit\end{tabular} & \begin{tabular}[c]{@{}c@{}}1 ALU, 1 Vector Unit\\ 1 FP Mult, 1 Int Mult\end{tabular} & 1 Load Unit & 1 Load Unit & 1 Store Unit & 1 ALU, 1 Vector Unit & 1 ALU, 1 Branch Unit \\
\hline
I & Cedarview & \begin{tabular}[c]{@{}c@{}}1 ALU,  1 Load Unit\\ 1 Store Unit, 1 Vector Unit\\ 1 Int Mult, 1 Divider\end{tabular} & \begin{tabular}[c]{@{}c@{}}1 ALU, 1 Vector Unit\\ 1 FP Unit, 1 Branch Unit\end{tabular} & 1 Load Unit & 1 Store Unit & - & - & - \\
\hline
I & Jaguar & 1 ALU, 1 Vector Unit & 1 ALU, 1 Vector Unit & 1 FP Unit, 1 Int Mult & 1 FP Mult, 1 Divider & 1 Load Unit & 1 Store Unit & - \\
\hline
I & Artificial 2 & \begin{tabular}[c]{@{}c@{}}1 ALU, 1 Vector Unit\\ 1 FP Unit, 1 Int Mult\\ 1 Divider, 1 Branch Unit\end{tabular} & \begin{tabular}[c]{@{}c@{}}1 ALU, 1 Vector Unit\\ 1 FP Mult, 1 FP Unit\\ 1 Int Mult\end{tabular} & 1 Load Unit & 1 Load Unit & 1 Store Unit & 1 ALU, 1 Vector Unit & 1 ALU, 1 Branch Unit \\
\hline
I & Artificial 3 & \begin{tabular}[c]{@{}c@{}}1 ALU, 1 Vector Unit\\ 1 FP Unit, 1 Int Mult\\ 1 Divider, 1 Branch Unit\end{tabular} & \begin{tabular}[c]{@{}c@{}}1 ALU, 1 Vector Unit\\ 1 FP Mult, 1 FP Unit\\ 1 Int Mult\end{tabular} & 1 Load Unit & 1 Load Unit & 1 Store Unit & 1 ALU, 1 Vector Unit & 1 ALU, 1 Branch Unit \\
\hline
I & Artificial 4 & \begin{tabular}[c]{@{}c@{}}1 ALU, 1 FP Mult\\ 1 FP unit, 1 Int Vector\\ 1 Int Mult, 1 Divider\\ 1 Branch Unit\end{tabular} & \begin{tabular}[c]{@{}c@{}}1 ALU, 1 Vector Unit\\ 1 FP Mult, 1 Int Mult\end{tabular} & 1 Load Unit & 1 Load Unit & 1 Store Unit & 1 ALU, 1 Vector Unit & 1 ALU, 1 Branch Unit \\
\hline
I & Artificial 6 & \begin{tabular}[c]{@{}c@{}}1 ALU, 1 Vector Unit\\ 1 FP Unit, 1 Int Mult\\ 1 Divider, 1 Branch Unit\end{tabular} & \begin{tabular}[c]{@{}c@{}}1 ALU, 1 Vector Unit\\ 1 FP Mult, 1 FP Unit\\ 1 Int Mult\end{tabular} & 1 Load Unit & 1 Load Unit & 1 Store Unit & 1 ALU, 1 Vector Unit & 1 ALU, 1 Branch Unit \\
\hline
I & Artificial 7 & 1 Load Unit, 1 Store Unit & 1 ALU, 1 Integer Mult & 1 ALU, 1 Branch Unit & 1 FP Mult, 1 Divider & 1 FP Unit & - & - \\
\hline
I & Artificial 10 & \begin{tabular}[c]{@{}c@{}}1 ALU,  1 Load Unit\\ 1 Store Unit, 1 Vector Unit\\ 1 Int Mult, 1 Divider\end{tabular} & \begin{tabular}[c]{@{}c@{}}1 ALU, 1 Vector Unit\\ 1 FP Unit, 1 Branch Unit\end{tabular} & 1 Load Unit & 1 Store Unit & - & - & - \\
\hline
I & Artificial 11 & \begin{tabular}[c]{@{}c@{}}1 ALU, 1 Load Unit\\ 1 Store Unit, 1 Vector Unit\\ 1 Int Mult, 1 Divider\end{tabular} & \begin{tabular}[c]{@{}c@{}}1 ALU, 1 Vector Unit\\ 1 FP Unit, 1 Branch Unit\end{tabular} & 1 Load Unit & 1 Store Unit & - & - & - \\
\hline
II & Ivybridge & \begin{tabular}[c]{@{}c@{}}1 ALU, 1 Vector Unit\\ 1 FP Mult, 1 Divider\end{tabular} & \begin{tabular}[c]{@{}c@{}}1 ALU, 1 Vector Unit\\ 1 Int Mult, 1 FP Unit\end{tabular} & 1 Load Unit & 1 Load Unit & 1 Store Unit & \begin{tabular}[c]{@{}c@{}}1 ALU, 1 Vector Unit\\ 1 Branch Unit, 1 FP Unit\end{tabular} & - \\
\hline
II & Artificial 0 & \begin{tabular}[c]{@{}c@{}}1 ALU, 1 FP Mult\\ 1 FP unit, 1 Int Vector\\ 1 Int Mult, 1 Divider\\ 1 Branch Unit\end{tabular} & \begin{tabular}[c]{@{}c@{}}1 ALU, 1 Vector Unit\\ 1 FP Mult, 1 Int Mult\end{tabular} & 1 Load Unit & 1 Load Unit & 1 Store Unit & 1 ALU, 1 Vector Unit & 1 ALU, 1 Branch Unit \\
\hline
II & Artificial 9 & \begin{tabular}[c]{@{}c@{}}1 ALU, 1 Vector Unit\\ 1 Int Mult\end{tabular} & 1 ALU, 1 Vector Unit & 1 ALU, 1 Vector Unit & 1 Load Unit & 1 Store Unit & 1 FP Unit & 1 FP Unit \\
\hline
III & Artificial 1 & \begin{tabular}[c]{@{}c@{}}1 ALU, 1 Vector Unit\\ 1 Int Mult\end{tabular} & 1 ALU, 1 Vector Unit & 1 ALU, 1 Vector Unit & 1 Load Unit & 1 Store Unit & 1 FP Unit & 1 FP Unit \\
\hline
III & Artificial 5 & \begin{tabular}[c]{@{}c@{}}1 ALU, 1 Vector Unit\\ 1 Int Mult\end{tabular} & 1 ALU, 1 Vector Unit & 1 ALU, 1 Vector Unit & 1 Load Unit & 1 Store Unit & 1 FP Unit & 1 FP Unit \\
\hline
III & Artificial 8 & \begin{tabular}[c]{@{}c@{}}1 ALU, 1 Vector Unit\\ 1 Int Mult\end{tabular} & 1 ALU, 1 Vector Unit & 1 ALU, 1 Vector Unit & 1 Load Unit & 1 Store Unit & 1 FP Unit & 1 FP Unit \\
\hline
IV & K8 & \begin{tabular}[c]{@{}c@{}}1 ALU, 1 Vector Unit\\ 1 Int Mult\end{tabular} & 1 ALU, 1 Vector Unit & 1 ALU, 1 Vector Unit & 1 Load Unit & 1 Store Unit & 1 FP Unit & 1 FP Unit \\
\hline
IV & K10 & \begin{tabular}[c]{@{}c@{}}1 ALU, 1 Vector Unit\\ 1 Int Mult\end{tabular} & 1 ALU, 1 Vector Unit & 1 ALU, 1 Vector Unit & 1 Load Unit & 1 Store Unit & 1 FP Unit & 1 FP Unit \\
\hline
IV & Silvermont & 1 Load Unit, 1 Store Unit & 1 ALU, 1 Integer Mult & 1 ALU, 1 Branch Unit & 1 FP Mult, 1 Divider & 1 FP Unit & - & - \\
\hline
IV & Skylake & \begin{tabular}[c]{@{}c@{}}1 ALU, 1 Vector Unit\\ 1 FP Unit, 1 Int Mult\\ 1 Divider, 1 Branch Unit\end{tabular} & \begin{tabular}[c]{@{}c@{}}1 ALU, 1 Vector Unit\\ 1 FP Mult, 1 FP Unit\\ 1 Int Mult\end{tabular} & 1 Load Unit & 1 Load Unit & 1 Store Unit & 1 ALU, 1 Vector Unit & 1 ALU, 1 Branch Unit \\
\hline
\end{tabular}%
}
\end{table*}
\end{scriptsize}

\subsection{Bug Development}
\label{subsec:bug_setup}

To the best of our knowledge, gem5 is a performance bug-free
simulator. To achieve wide bug representation
we reviewed errata of commercial processors,
consulted with industry experts and tried to cover
as many units as possible.
%
Ultimately, 14 basic types of bugs are
developed and are summarized as follows.  

\begin{compactenum}
\itemsep0em 
\item Serialize X: Every instruction with opcode of X is
  marked as a serialized instruction. This causes all future
  instructions (according to program order) to be stalled until the
  instruction with the bug has been issued.
    
\item Issue X only if oldest: Instructions whose opcode is X
  will only be retired from the instruction queue when it has become
  the oldest instruction there. This bug is similar to ``POPCNT
  Instruction may take longer to execute than expected" found on Intel
  Xeon Processors~\cite{intel_xeon_errata}.
  
\item If X is oldest, issue only X: When an instruction whose opcode
  is X becomes the oldest in the instruction queue, only that
  instruction will be issued, even though other instructions might
  also be ready to be issued and the computation resources allow it.
    
\item If X depends on another instruction Y, delay T
  cycles. 
    
\item If less than N slots available in instruction queue, delay T
  cycles. 
    
\item If less than N slots available in re-order buffer, delay T
  cycles. 
    
\item If mispredicted branch, delay T
  cycles. 
    
\item If N stores to cache line, delay T cycles: After N stores have
  been executed to the same cache line, upcoming stores to the same
  line will be delayed by T cycles. This is a
  variation of ``Store gathering/merging performance issue" found
  on NXP MPC7448 RISC processor \cite{nxp_7448_errata}.
    
\item After N stores to the same register, delay T
  cycles. 
  This bug is inspired by ``GPMC may Stall After 256 Write Accesses in
  NAND\_DATA, NAND\_COMMAND, or NAND\_ADDRESS Registers" found on TI
  AM3517, AM3505 ARM processor~\cite{TI_am3517_errate}, 
  we generalized it for any physical register, and in our
  case, the instruction is only stalled for a few cycles, as opposed
  to the actual bug, where the processor hanged. Another variation of this bug is implemented where the delay is applied once every N stores, instead of every store after the N-th.
  
\item L2 cache latency is increased by T
  cycles. 
  This bug is inspired by the ``L2 latency performance issue" on the
  NXP MPC7448 RISC processor~\cite{nxp_7448_errata}.
  
\item Available registers reduced by
  N. 
    
\item If branch longer than N bytes, delay T
  cycles. 
    
\item If X uses register R, delay T
  cycles. 
  This bug is a variation of the ``POPA/POPAD Instruction Malfunction"
  found on Intel 386 DX \cite{intel_386_errata}.
    
\item Branch predictor's table index function issue, reducing effective table size by N entries.

\end{compactenum}

\begin{figure}[!hbt]
    \centering
    \includegraphics[width = 0.33\textwidth]{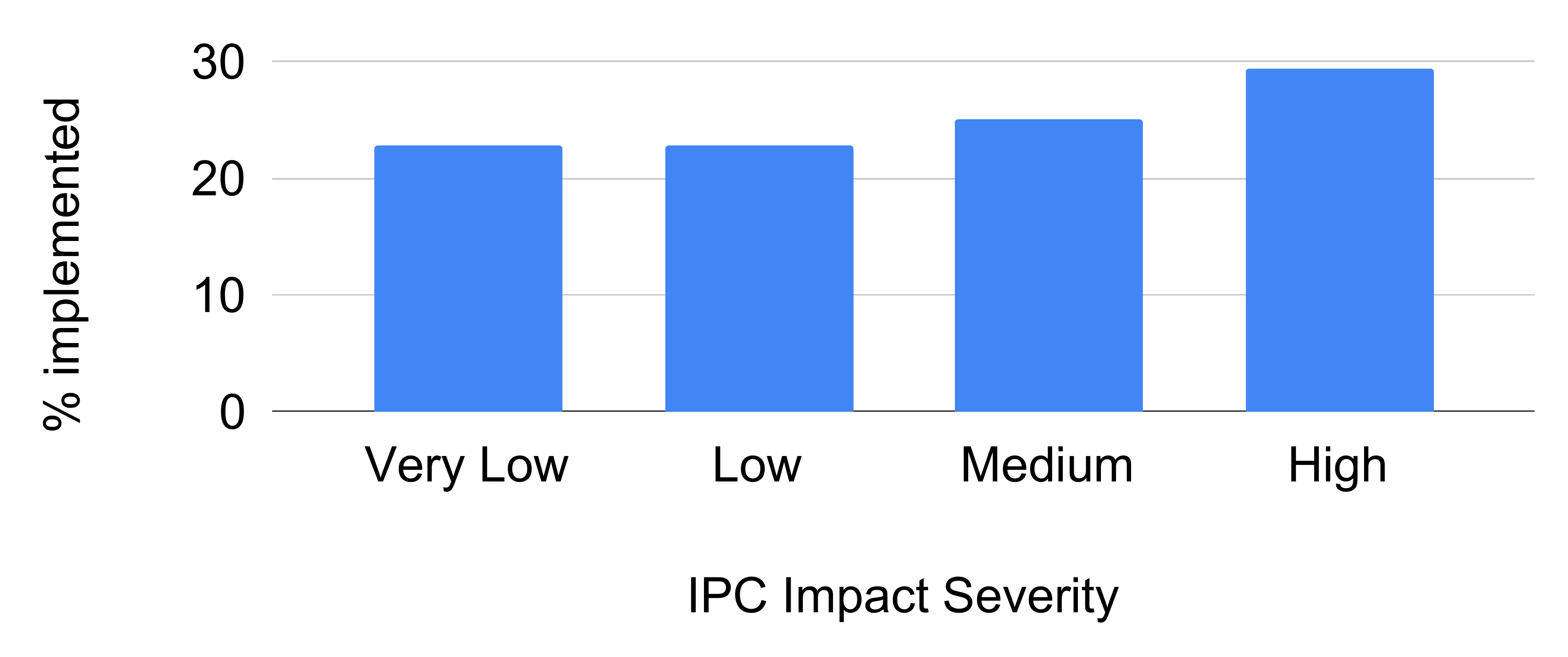}
    \caption{Distribution of bug severity.}
    \label{fig:bug_dist}
\end{figure}

For each of these bug types, multiple variants are implemented by
changing \emph{X, Y, N, R} and \emph{T} respectively.  The bugs are
grouped into four categories according to their impact on IPC:
\textit{High} means an average IPC degradation (across the used
SPEC CPU2006 applications) of 10\% or greater. A \textit{Medium} impact
means the average IPC is degraded between 5\% and 10\%. A \textit{Low}
impact is between 1\% and 5\% and a \textit{Very-Low} impact is less
than 1\%. Figure \ref{fig:bug_dist} shows the distribution of the
severity of average IPC impact for the implemented bugs.

\subsection{\hlcyan{Bug Detection in Memory
    Systems}}
\label{subsec:detection_mem}

We also examine the performance of the proposed methodology on areas
outside of the processor core, specifically on the processor cache
memory system.  Overall the methodology remains unchanged, but there
are some minor differences vs. the evaluation of our technique on
processor cores.  Here we use the ChampSim~\cite{champsim} simulator
instead of gem5~\cite{gem5}, because it provides a more
detailed memory system simulation with relatively short simulation time.
ChampSim was developed for rapid evaluation of
microarchitectural techniques in the processor cache hierarchy. 

The probes we use for this evaluation correspond to 22 SimPoints from
seven applications on the SPEC CPU2006 \cite{spec2006} benchmark
suite. The
emulated architectures are Intel's Broadwell, Haswell, Skylake,
Sandybridge, Ivybridge, Nehalem and AMD's K10 and Ryzen7, as well as four artificial architectures.
The developed bugs are summarized as follows.

\begin{compactenum}
\item When a cache block is accessed, the age counter for the replacement policy is not updated.
\item During a cache eviction, the policy evicts the most recently used block, instead of the least recently used.
\item After N load misses on L1D cache, the read operation is delayed by T clock cycles. Another variant of this bug was implemented for L2 cache.
\item Signature Path Prefetcher (SPP) \cite{kim2016spp} signatures are reset, making the prefetcher use the wrong address.
\item On lookahead prefetching, the path with the least confidence is selected.
\item Some prefetches are incorrectly marked as executed. This bug was found in the SPP \cite{kim2016spp} prefetching method.
\end{compactenum}

Since IPC might be affected by many other factors outside the memory
system, we evaluate our methodology by using Average Memory Access
Time (AMAT) as the target metric for the performance models.
\shepherd{Given the differences between simulators, architectures and bugs, the rules
for stage 2 were slightly modified for this evaluation, however, the overall 
methodology remains unchanged.}

\section{Evaluation}
\label{sec:results}

\hlcyan{Since the main focus of our work is on the processor core,
  sections} \ref{subsec_exp_stage1} to
\ref{subsec_exp_training_uarchs} \hlcyan{examine performance bugs in
  that unit. Section} \ref{subsec_exp_mem} \hlcyan{expands
  examination to memory system bugs.}




\subsection{Machine Learning-Based IPC Modeling}
\label{subsec_exp_stage1}

In this section, we show the results for the first stage of our
methodology, IPC modelling.  We evaluated the performance of several
machine learning methods, as discussed in
Section~\ref{subsec:method_stg_1}, and different variations for each
of them.  The MLP, CNN and LSTM networks are implemented using Keras
library~\cite{chollet2015keras}. In training all these models, Mean
Squared Error (MSE) is used as the loss function and
Adam~\cite{kingma2015adam} is employed as the optimizer.  A gradient
clipping of $0.01$ is enforced to avoid the gradient explosion issue,
commonly seen in training recurrent networks.  The gradient boosted
trees are implemented via the XGBoost
library~\cite{chen2016xgboost}. Lasso is implemented using
  Scikit-Learn~\cite{scikit-learn}.

The total (all probes) training and inference runtime, as well as
inference error (Eq.~\ref{eq:error}) for each IPC modelling technique
are shown in Table~\ref{tab:ipc_mse_auc}. The name of each neural
network-based method (LSTM, CNN and MLP), is prefixed with a
number indicating the number of hidden layers, and postfixed with
the number of neurons in each hidden layer. The
postfix number for a GBT method is the number of trees. Runtime
results are measured on an Intel Xeon E5-2697A V4 processor with
frequency of 2.6GHz and 512GB memory\footnote{This machine has no GPU,
  and GPU acceleration could significantly reduce these
  runtimes.}. 
The total
simulation time for detecting bugs in one new microarchitecture takes
about 6 hours if the simulations are not executed in parallel.

\begin{scriptsize}
\begin{table}[!htb]
\centering
\caption{IPC modelling runtime and error statistics.}
\resizebox{0.48\textwidth}{!}{
\label{tab:ipc_mse_auc}
\begin{tabular}{c|cc|cccc}
\hline
\multicolumn{1}{l|}{\textbf{}} & \multicolumn{2}{c|}{\textbf{Runtime}}  & \multicolumn{4}{c}{\textbf{Inference Error (Equation~\eqref{eq:error}) } } \\
\textbf{ML Model} & \textbf{Training} & \textbf{Inference} & \textbf{Average} & \textbf{Std. Dev.} & \textbf{Median} & \textbf{90th Perc.} \\ \hline
\hlcyan{\textbf{Lasso}} & \hlcyan{0h 8m} & \hlcyan{3m 06s} & \hlcyan{10.2749} & \hlcyan{2.1598} & \hlcyan{10.3995} & \hlcyan{13.1740} \\
\textbf{1-LSTM-150} & 3h 22m & 16m 21s & $2.87\times10^{7}$ & $2.89\times10^{8}$ & 2.8082 & 5.4062 \\
\textbf{1-LSTM-250} & 4h 17m & 17m 28s & $4.94\times10^{23}$ & $6.71\times10^{24}$ & 2.8364 & 5.4582 \\
\textbf{1-LSTM-500} & 3h 39m & 24m 18s & $1.2\times10^{6}$ & $1.63\times10^{7}$ & 2.9193 & 4.7848 \\
\textbf{4-LSTM-150} & 4h 39m & 54m 33s & $5.81\times10^{7}$ & $5.87\times10^{8}$ & 5.5581 & 13.2453 \\
\textbf{4-LSTM-500} & 2h 54m & 71m 20s & $1.09\times10^{3}$ & $1.35\times10^{4}$ & 4.3551 & 11.7138 \\
\textbf{1-CNN-150} & 0h 19m & 08m 16s & 5.3959 & 5.3458 & 3.4171 & 17.1287 \\
\textbf{4-CNN-150} & 0h 33m & 15m 53s & 5.7568 & 5.4779 & 3.5433 & 16.2600 \\
\textbf{1-MLP-500} & 1h 38m & 08m 39s & 2.2506 & 1.5636 & 1.8200 & 4.2648 \\ 
\textbf{1-MLP-2500} & 1h 51m & 07m 01s & 2.1298 & 1.6228 & 1.6589 & 4.4684 \\ 
\textbf{4-MLP-500} & 3h 19m & 10m 05s & 9.8390 & 66.6996 & 4.1207 & 6.6891 \\ 
\textbf{GBT-150} & 0h 30m & 5m 01s & 3.7928 & 2.1445 & 3.3095 & 6.3616 \\
\textbf{GBT-250} & 0h 38m & 4m 34s & 3.6181 & 2.0607 & 3.1275 & 6.1395 \\
\hline
\end{tabular}
}
\end{table}
\end{scriptsize}

\begin{figure*}[!hbt]
    \centering
    \begin{subfigure}[b]{0.32\textwidth}
     \includegraphics[height=3.75cm,width=0.85\textwidth]{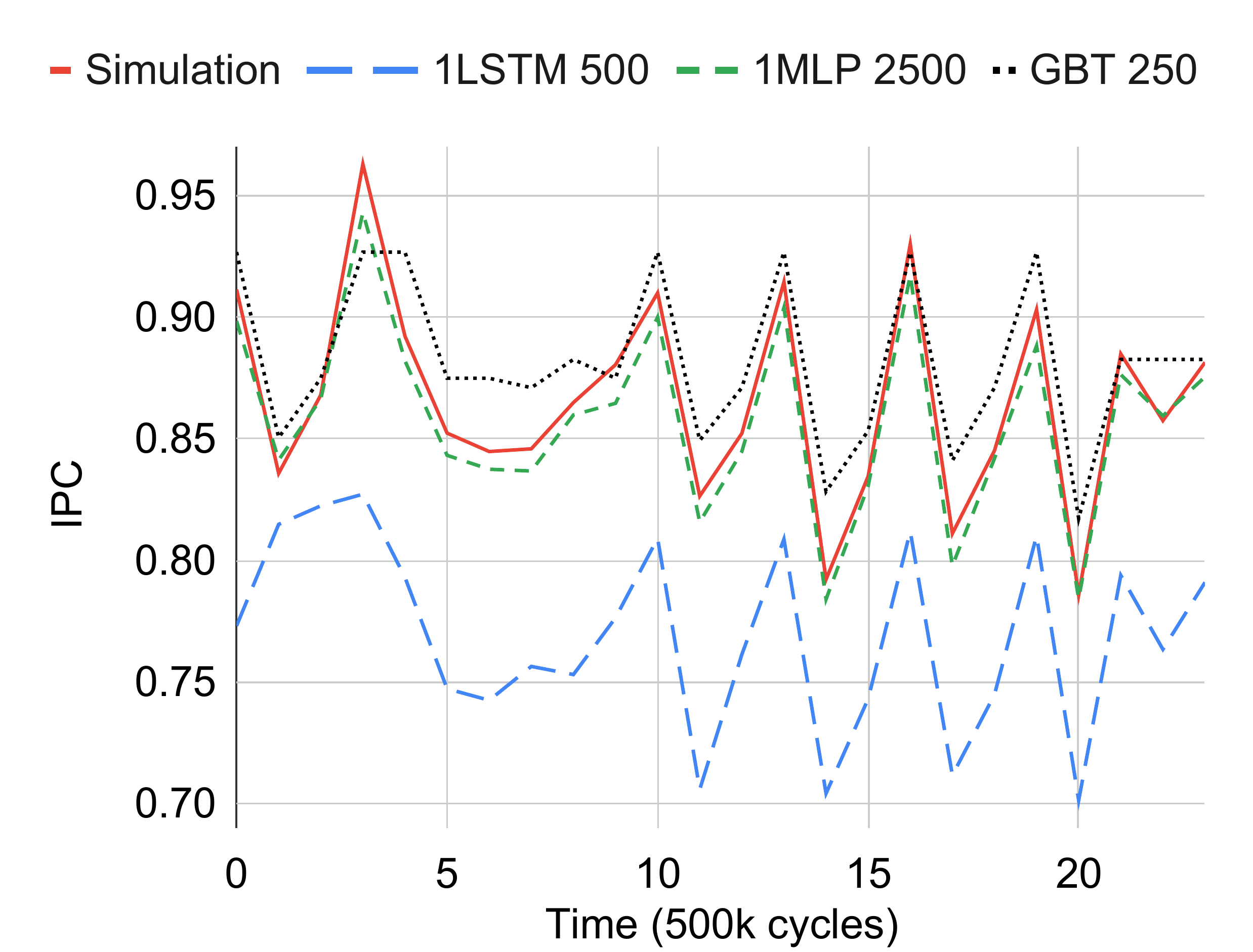}
        \caption{SimPoint \#12 of gcc}
        \label{fig:no_bug_trace_gcc_12}
    \end{subfigure}
    ~ 
    \begin{subfigure}[b]{0.32\textwidth}
     \includegraphics[height=3.75cm,width=0.85\textwidth]{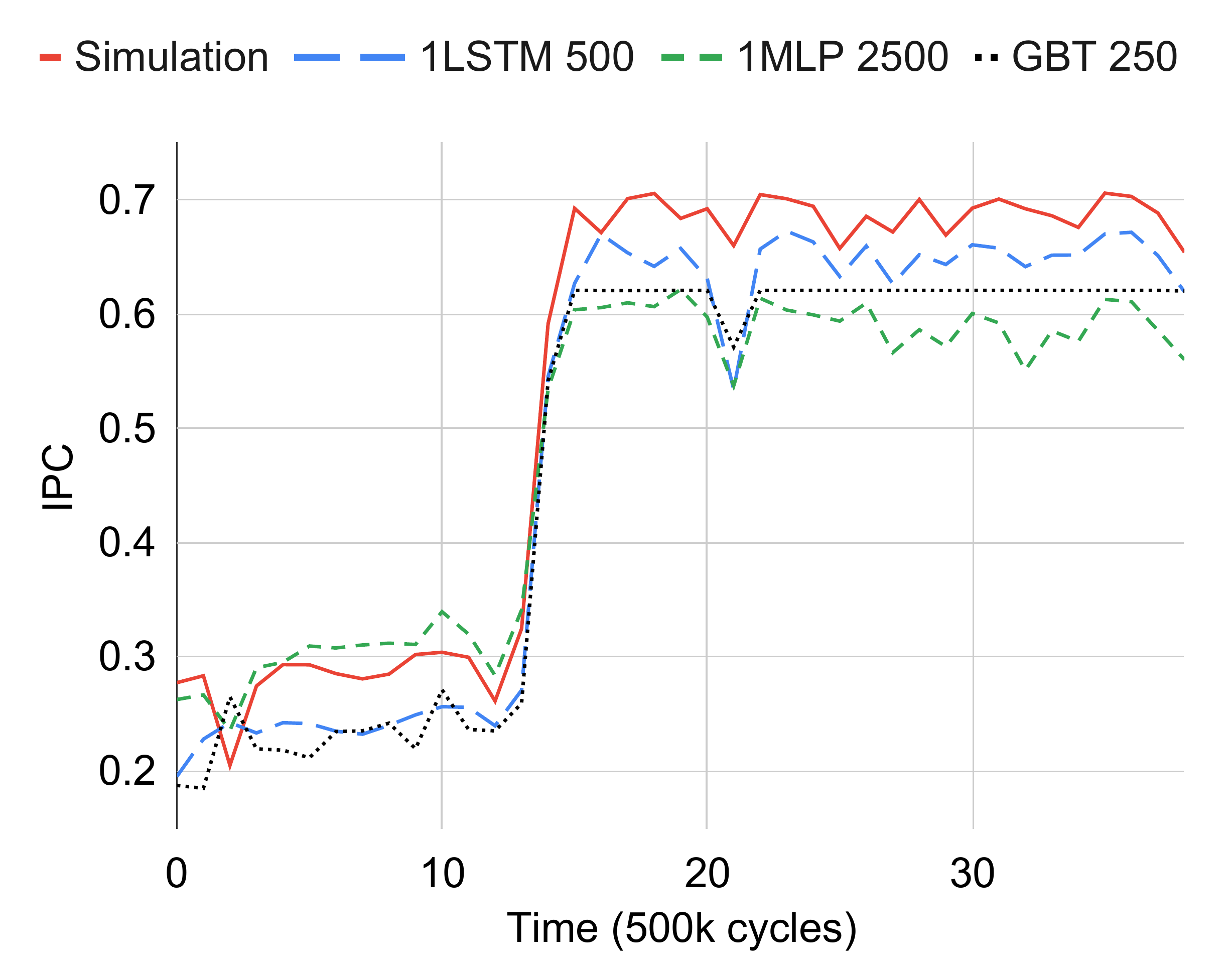}
       \caption{SimPoint \#16 of bzip2}
        \label{fig:no_bug_trace_bzip2_16}
    \end{subfigure}
    \begin{subfigure}[b]{0.32\textwidth}
     \includegraphics[height=3.75cm,width=0.85\textwidth]{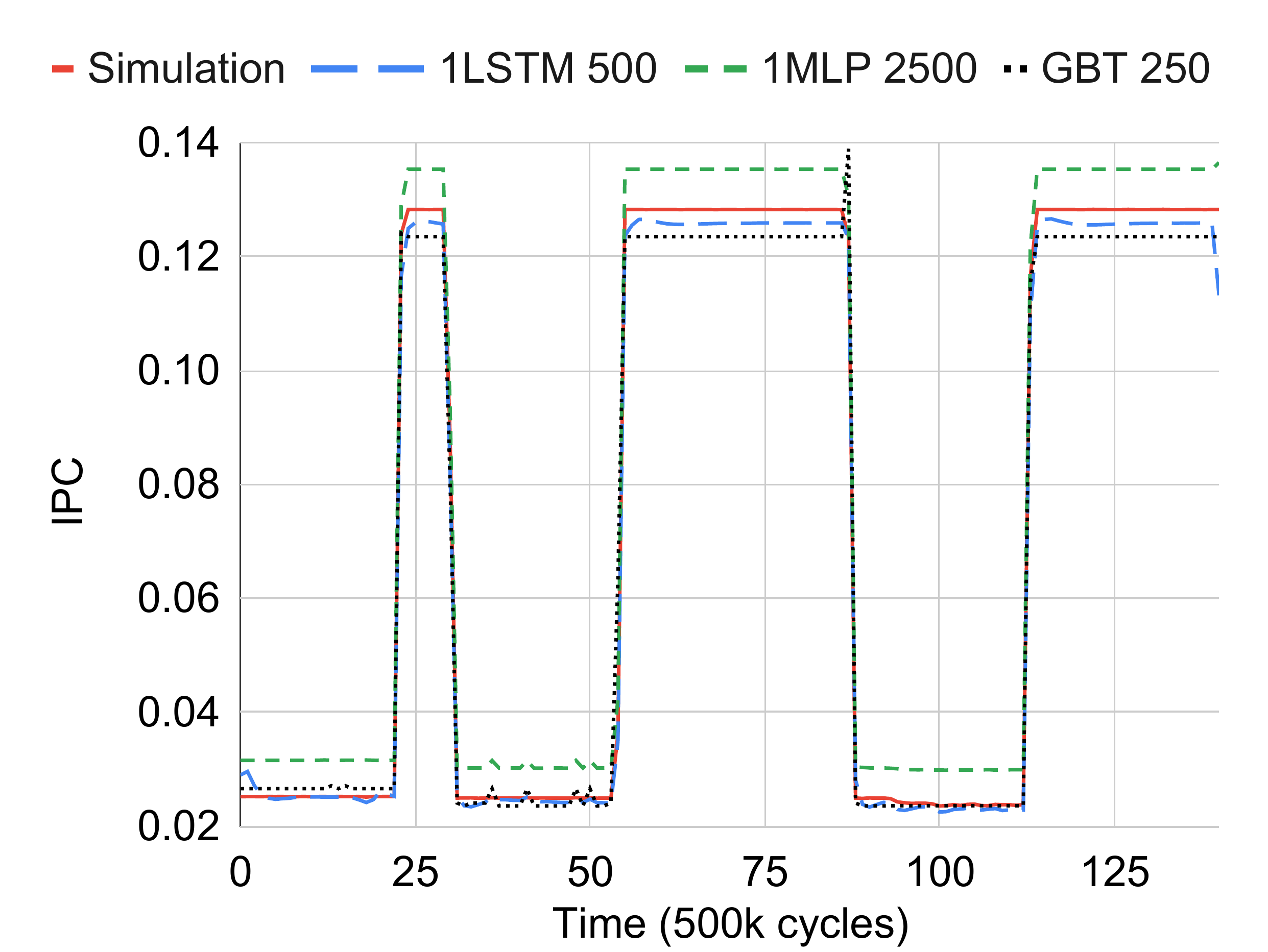}
       \caption{SimPoint \#1 of cactusADM}
        \label{fig:no_bug_trace_cactus_00}
    \end{subfigure}
        \vskip 0.5ex
    \caption{Examples of ML-based IPC inference and simulated IPC on bug-free microarchitectures.}
    \vskip 1ex
    \label{fig:fig:no_bug_trace}
\end{figure*}

\begin{figure*}[!hbt]
    \centering
    \begin{subfigure}[b]{0.23\textwidth}
     \includegraphics[width=0.98\textwidth]{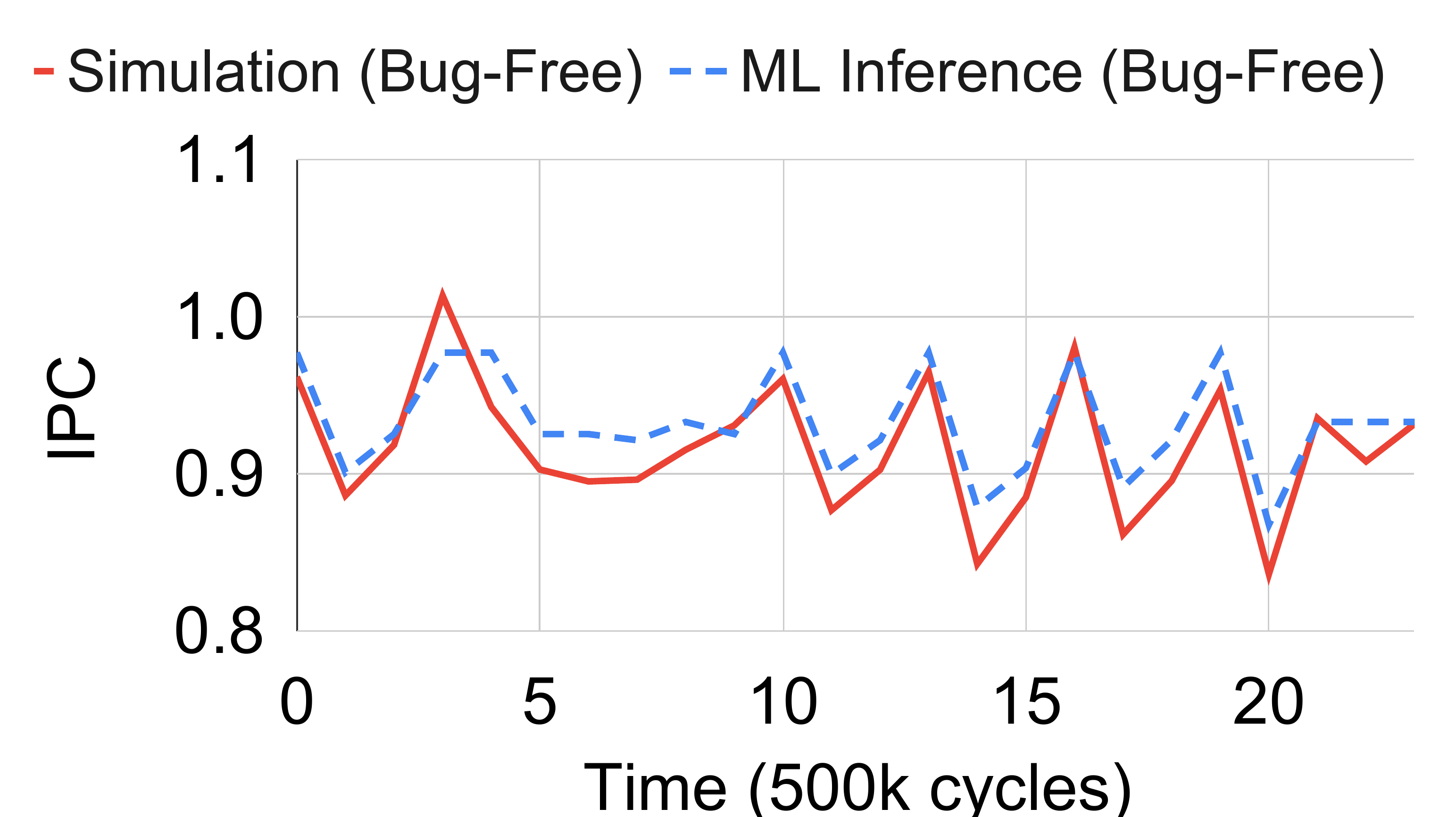}
        \caption{Bug-Free SimPoint \#12 in gcc}
        \label{fig:if_oldest_issue_xor_nobug}
    \end{subfigure}
    ~ 
    \begin{subfigure}[b]{0.23\textwidth}
     \includegraphics[width=0.98\textwidth]{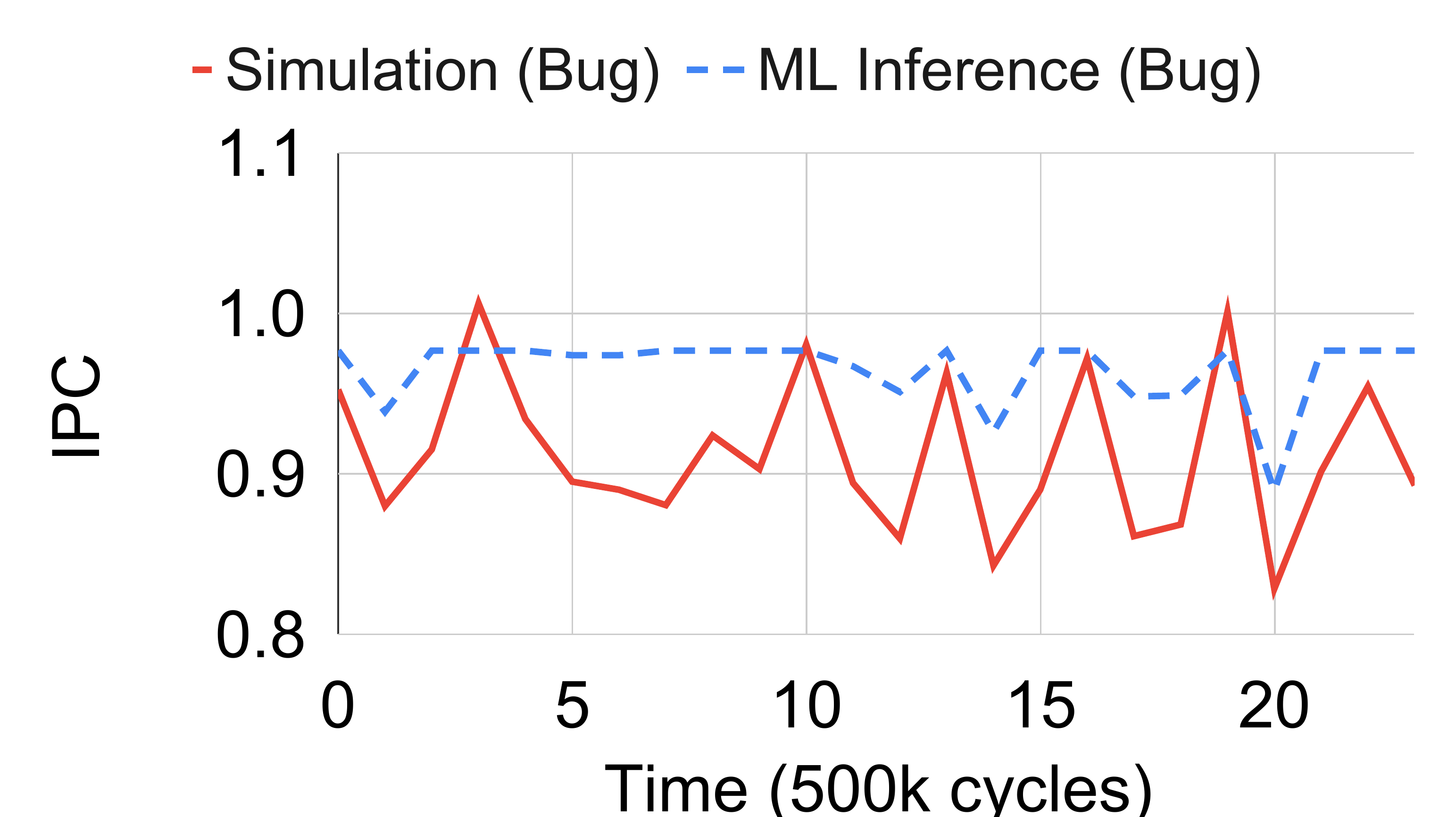}
       \caption{Bug SimPoint \#12 in gcc}
        \label{fig:if_oldest_issue_xor_bug}
    \end{subfigure}
    ~ 
    \begin{subfigure}[b]{0.25\textwidth}
     \includegraphics[width=0.93\textwidth]{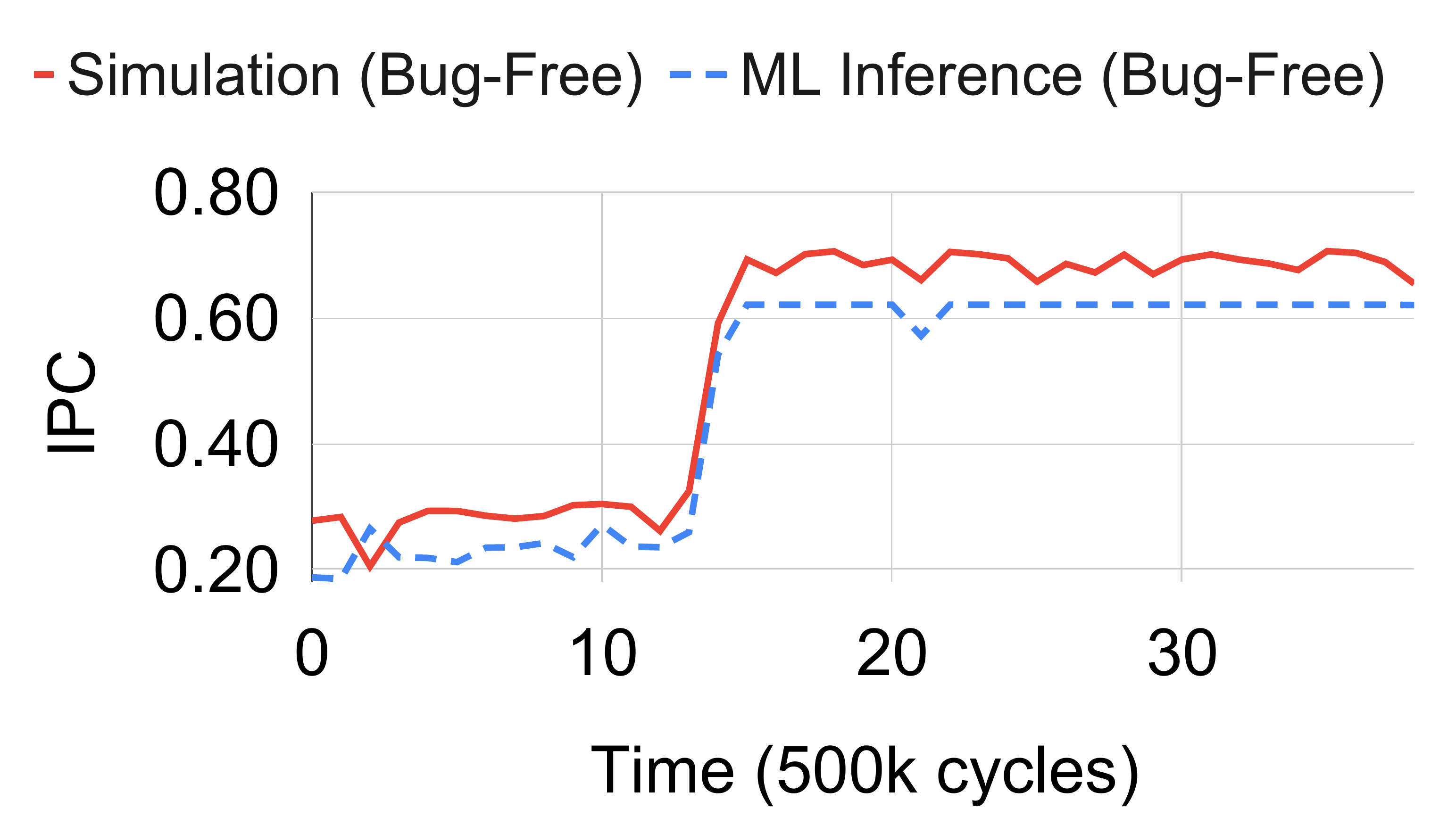}
       \caption{Bug-Free SimPoint \#16 in bzip2}
        \label{fig:serialized_sub_nobug}
    \end{subfigure}
    ~ 
    \begin{subfigure}[b]{0.23\textwidth}
     \includegraphics[width=0.98\textwidth]{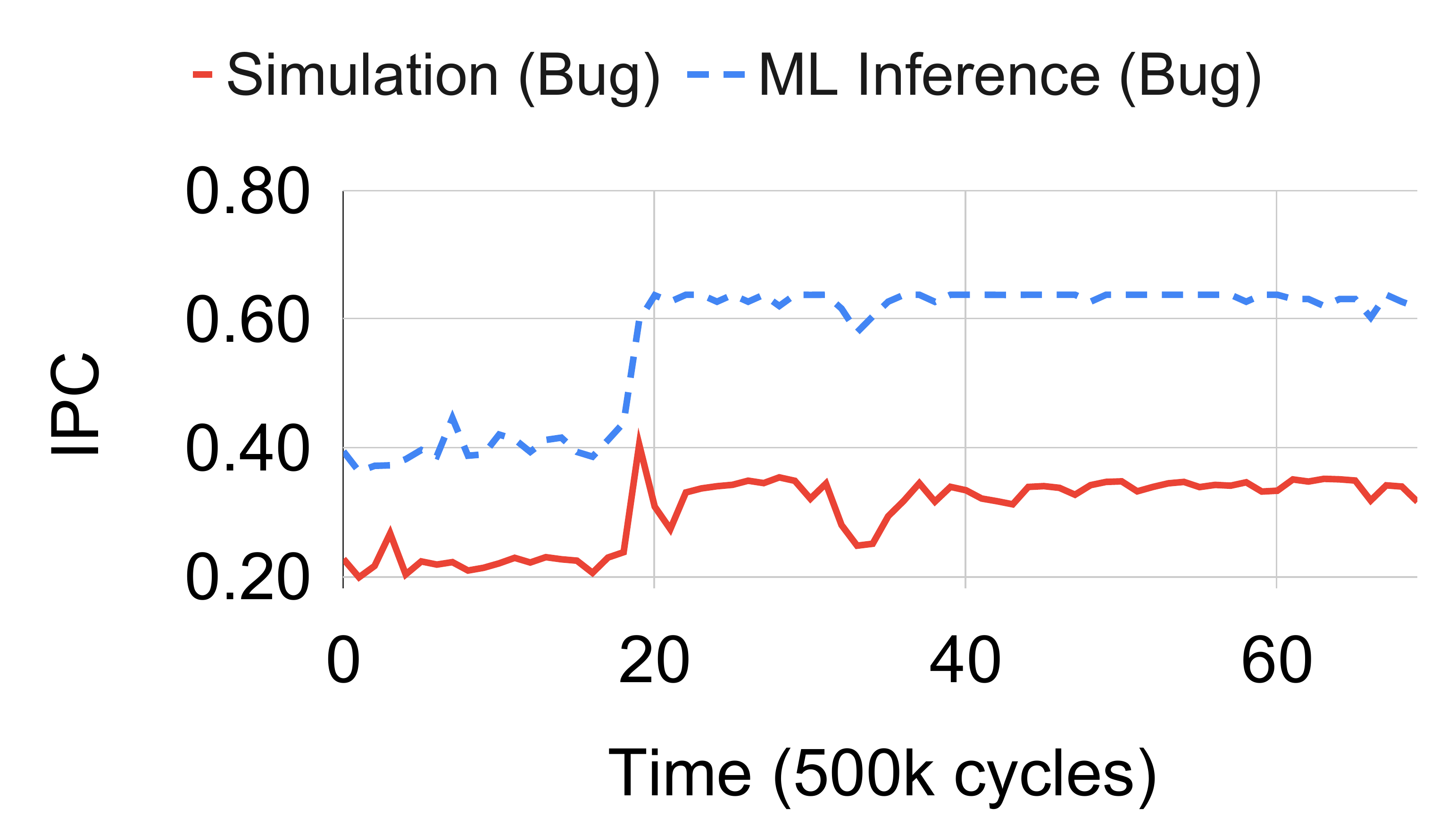}
       \caption{Bug SimPoint \#16 in bzip2}
        \label{fig:serialized_sub_bug}
    \end{subfigure}
    \vskip 0.5ex
    \caption{IPC estimations by GBT-250: comparisons between microarchitectures with and without bugs.}
    \label{fig:bug_contrast}
\end{figure*}

The right four columns of Table~\ref{tab:ipc_mse_auc} summarize the
inference errors defined in Equation~\eqref{eq:error}. Please note
IPC inference is a regression task, not classification.  Thus,
classification metrics, such as true positive and false positive
rates, do not apply.  The results are averaged across all probes
for bug-free designs of microarchitectures in Set IV.  Median
errors of different ML engines are close to each other. LSTM has huge
variances and average errors due to a few non-convergent outliers.
Although input features are a time series, where LSTM should
excel, the actual LSTM errors here are no better than non-recurrent
networks and sometimes much worse.  There are two reasons for this.
First, the time step size we chose is large enough such that the
history effect has already been well counted in one time step and thus
the recurrence in LSTM becomes unnecessary. Second, LSTM is well-known
to be difficult to train and the difficulty is confirmed by those
outlier cases.  In stage 2 of our methodology, LSTM results with huge
errors are not used.

Figure~\ref{fig:fig:no_bug_trace} displays the IPC time series for
three SimPoints (chosen to represent varied behavior), on
the Skylake microarchitecture, where the red solid lines indicate the
measured (simulated) IPCs, while the dotted/dashed lines are IPC
inferences of the different ML engines. In general, the ML models
trace IPCs very well.  LSTM has relatively large errors in
Figure~\ref{fig:no_bug_trace_gcc_12}, yet it still shows strong
correlation with the simulated IPCs.  In all three cases the results
confirm the effectiveness of the machine learning models across
various scenarios.

Although IPC inference accuracy for bug-free designs is
important, the inference error difference between cases with and
without bugs matters even more. Such difference is illustrated for two
SimPoints in Figure~\ref{fig:bug_contrast}.  In Figure
\ref{fig:if_oldest_issue_xor_nobug}, where the microarchitecture is
bug-free, GBT-250 estimates IPCs very accurately.  When there is a bug,
however, as shown in Figure~\ref{fig:if_oldest_issue_xor_bug}, the
inference errors drastically increase.  The same discrepancy is also
exhibited between another bug-free design,
Figure~\ref{fig:serialized_sub_nobug}, and buggy design,
Figure~\ref{fig:serialized_sub_bug}.  Both the examples demonstrate
that a significant loss of accuracy implies bug existence.





\subsection{Bug Detection}


In this section, we show the evaluation results for our stage 2
classification given the IPC inference errors from stage 1. The
evaluation includes comparisons with the na\"ive single-stage baseline
approach described in Section~\ref{sec:baseline}.  The probe designs
of the baseline are the same as our proposed methodology.  Here we
only use the GBT-250 engine which has the best single-stage results.
Its model features include simulated IPCs in addition to data from the
selected counters and microarchitectural design parameters. However,
it uses a single value for each feature aggregated from an entire
SimPoint instead of the time series.

The bug detection efficacy is evaluated by the metrics below.
\begin{equation}
    \label{eq:metrics}
    \textrm{FPR} = \frac{\textrm{FP}}{\textrm{N}}, ~~ \textrm{Precision} = \frac{\textrm{TP}}{\textrm{TP}+\textrm{FP}}, ~~
    \textrm{TPR} = \frac{\textrm{TP}}{\textrm{P}}
\end{equation}
where N and P are the number of real negative (no bug) and positive
(bug) cases, respectively. FP (False Positive) indicates the number of
cases that are incorrectly classified as having bugs.  TP (True Positive)
is the number of cases that are correctly classified as having
bugs. Additionally, ROC (Receiver Operating Characteristic) AUC (Area
Under Curve) is evaluated.  ROC shows the trade-off between TPR and
FPR. ROC AUC value varies between 0 and 1. A random guess would result
in ROC AUC of 0.5 and a perfect classifier can achieve 1.  Accuracy,
another common metric, is not employed here as the number of negative
cases is too small, i.e., the testcases are imbalanced.

\begin{table}[!htb]
\centering
\caption{Bug detection results. Here, Bug 1 is ``If XOR is oldest in
  IQ, issue only XOR'' and Bug 2 is ``If ADD uses register 0, delay 10
  cycles''.}
\resizebox{0.475\textwidth}{!}{
\label{tab:class_sch_2}

\begin{tabular}{c|c|c|c|c|c|cccc}
\hline
\multicolumn{1}{l|}{} & \multicolumn{1}{l|}{} & \multicolumn{1}{l|}{} & \multicolumn{1}{l|}{} & \multicolumn{1}{l|}{} & \multicolumn{1}{l|}{} & \multicolumn{4}{c}{\textbf{TPR for different bug categories}} \\
\textbf{\begin{tabular}[c]{@{}c@{}}\hlcyan{Bugs in presumed}\\ \hlcyan{bug-free training}\end{tabular}} & \textbf{\begin{tabular}[c]{@{}c@{}}Stage 1\\ ML Model\end{tabular}}  & \textbf{FPR} & \textbf{TPR} & \textbf{\begin{tabular}[c]{@{}c@{}}ROC\\ AUC\end{tabular}} & \textbf{Precision} & \textbf{High} & \textbf{Medium} & \textbf{Low} & \textbf{Very Low} \\ \hline
\hlcyan{No Bug} & \textbf{Single-stage baseline} & 0.00 & 0.75 & 0.87 & 1.00 & 1.00 & 0.71 & 0.74 & 0.41 \\ \hline 
\hlcyan{No Bug} & \hlcyan{\textbf{Lasso}} & \hlcyan{0.20}  & \hlcyan{0.39} & \hlcyan{0.57} & \hlcyan{0.58} & \hlcyan{0.74} & \hlcyan{0.14} & \hlcyan{0.17} & \hlcyan{0.23} \\
\hlcyan{No Bug} & \textbf{1-LSTM-150} & 0.43 & 0.68 & 0.83 & 0.78 & 0.98 & 0.71 & 0.37 & 0.56 \\
\hlcyan{No Bug} & \textbf{1-LSTM-250} & 0.34 & 0.67 & 0.80 & 0.81 & 0.98 & 0.79 & 0.54 & 0.33 \\
\hlcyan{No Bug} & \textbf{1-LSTM-500} & 0.25 & 0.68 & 0.80 & 0.86 & 1.00 & 0.93 & 0.41 & 0.51 \\
\hlcyan{No Bug} & \textbf{4-LSTM-150} & 0.00 & 0.45 & 0.73 & 1.00 & 0.89 & 0.71 & 0.15 & 0.05 \\
\hlcyan{No Bug} & \textbf{4-LSTM-500} & 0.25 & 0.56 & 0.80 & 0.75 & 0.94 & 0.57 & 0.27 & 0.33 \\
\hlcyan{No Bug} & \textbf{1-CNN-150}  & 0.75 & 0.68 & 0.71 & 0.66 & 0.89 & 0.71 & 0.46 & 0.62 \\
\hlcyan{No Bug} & \textbf{4-CNN-150}  & 0.34 & 0.53 & 0.63 & 0.66 & 0.87 & 0.57 & 0.27 & 0.31 \\
\hlcyan{No Bug} & \textbf{1-MLP-500} & 0.50 & 0.84 & 0.88 & 0.81 & 1.00 & 1.00 & 0.83 & 0.59 \\
\hlcyan{No Bug} & \textbf{1-MLP-2500} & 0.48 & 0.80 & 0.85 & 0.80 & 0.98 & 0.93 & 0.73 & 0.59 \\ 
\hlcyan{No Bug} & \textbf{4-MLP-500} & 0.43 & 0.77 & 0.77 & 0.81 & 0.96 & 0.86 & 0.63 & 0.62 \\
\hlcyan{No Bug} & \textbf{GBT-150} & 0.00  & 0.80 & 0.89 & 1.00 & 0.96 & 0.93 & 0.73 & 0.59 \\
\hlcyan{No Bug} & \textbf{GBT-250} & 0.00  & 0.84 & 0.90 & 1.00 & 1.00 & 0.93 & 0.69 & 0.75 \\
\hline
\hlcyan{Bug 1} & \hlcyan{\textbf{GBT-250}} & \hlcyan{0.15}  & \hlcyan{0.69} & \hlcyan{0.83} & \hlcyan{0.90} & \hlcyan{0.94} & \hlcyan{0.31} & \hlcyan{0.61} & \hlcyan{0.54} \\
\hlcyan{Bug 2} & \hlcyan{\textbf{GBT-250}} & \hlcyan{0.20}  & \hlcyan{0.74} & \hlcyan{0.84} & \hlcyan{0.88} & \hlcyan{0.98} & \hlcyan{0.43} & \hlcyan{0.63} & \hlcyan{0.60} \\
\hline 
\end{tabular}
}
\end{table}

In our testing scheme, we attempt to detect a bug in a new microarchitecture
that is not included in training data.  Moreover, the training data does not include any bug with the same type as the one in the testing data, i.e., the bug in the new microarchitecture is completely unseen in the training. The data organization is as follows.

\begin{figure}[!hbt]
    \centering
    \includegraphics[width = 0.36\textwidth]{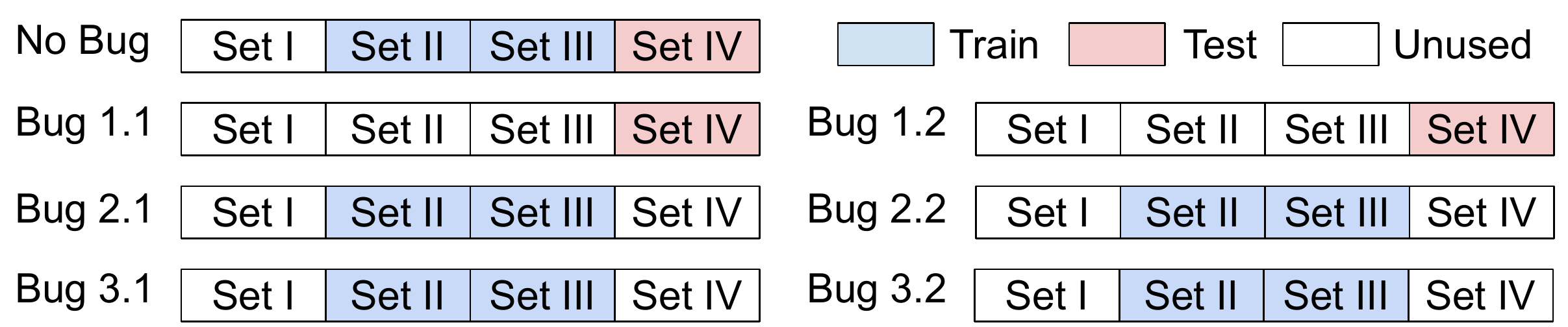}
    \caption{Example of training and testing data split.}
    \label{fig:new_arch_new_bug}
\end{figure}

\begin{itemize}
\item Training data:
  \begin{enumerate}
  \itemsep0em 

  \item Data with positive labels. IPC inference errors of all probes
    for microarchitectures in sets II and III with bug insertion. For
    each microarchitecture here, all types of bugs, except the one to
    be used in testing, are separately inserted. In each case, only
    one bug is inserted.
  \item Data with negative labels. IPC inference errors of all probes
    on bug-free versions of the microarchitectures in sets II and III.
\end{enumerate}
\item Testing data:
  \begin{enumerate}
  \itemsep0em 
  \item Designs with bugs. Microarchitectures in set IV with all
    variants of a bug type inserted and this bug type is not included
    in training data. In each case, only one bug is inserted.
  \item Bug-free versions of the microarchitectures in set IV.
\end{enumerate}
\end{itemize}
An example, which has 3 bug types (1, 2, 3) and 2 variants for each
type (\textit{e.g.} Bug 1.1 and Bug 1.2 are two variants of the same
bug type), is shown in Figure~\ref{fig:new_arch_new_bug}.

\hlcyan{Since previous architectures that are considered ``bug-free'' may
  actually have performance bugs, we also present results for the case when
  designs with a bug are presumed as ``bug-free'' and are used for
  training.}

The results are shown in Table~\ref{tab:class_sch_2}. Although the
same stage 2 classifier of our methodology is used, several different
ML engines are used in stage 1 and listed in the ``Stage 1 ML Model'' column.
\hlcyan{The leftmost column indicates whether bug-free or ``buggy''
  designs were used for training.}
  
  \hlcyan{When only bug-free designs are used,
  using GBT-250 in stage 1 produces the best result.  It can detect
  medium and high impact bugs ($> 5\%$ IPC impact) with 98.5\% true
  positive rate.  When low impact bugs ($> 1\%$ IPC impact) are
  additionally included, the true positive rate is still as high as
  $91.5\%$. The TPRs of GBT-250 beat the single-stage baseline in
  almost every bug category. It is also superior on ROC
  AUC. Meanwhile, it achieves 0 false-positive rate and 1.0
  precision.}

\hlcyan{The table also shows two cases where the models were trained using
  designs with a bug.  These two cases correspond to bugs
  with a low average IPC impact across the studied workloads. We
  included these as representative cases, and argue that bugs with
  higher IPC impact will most likely be caught during post-silicon
  validation of previous designs and will be fixed by the time a new
  microarchitecture is being developed. To evaluate these, we use
  GBT-250 model for stage 1 of the methodology as it was best
  performing.  As expected, these results show some degradation in
  detection. However, GBT-250 is still able to detect around 70\% of
  the bugs, while incurring a few false-positives.}



\begin{figure}[!hbt]
    \centering
    \includegraphics[width = 0.32\textwidth]{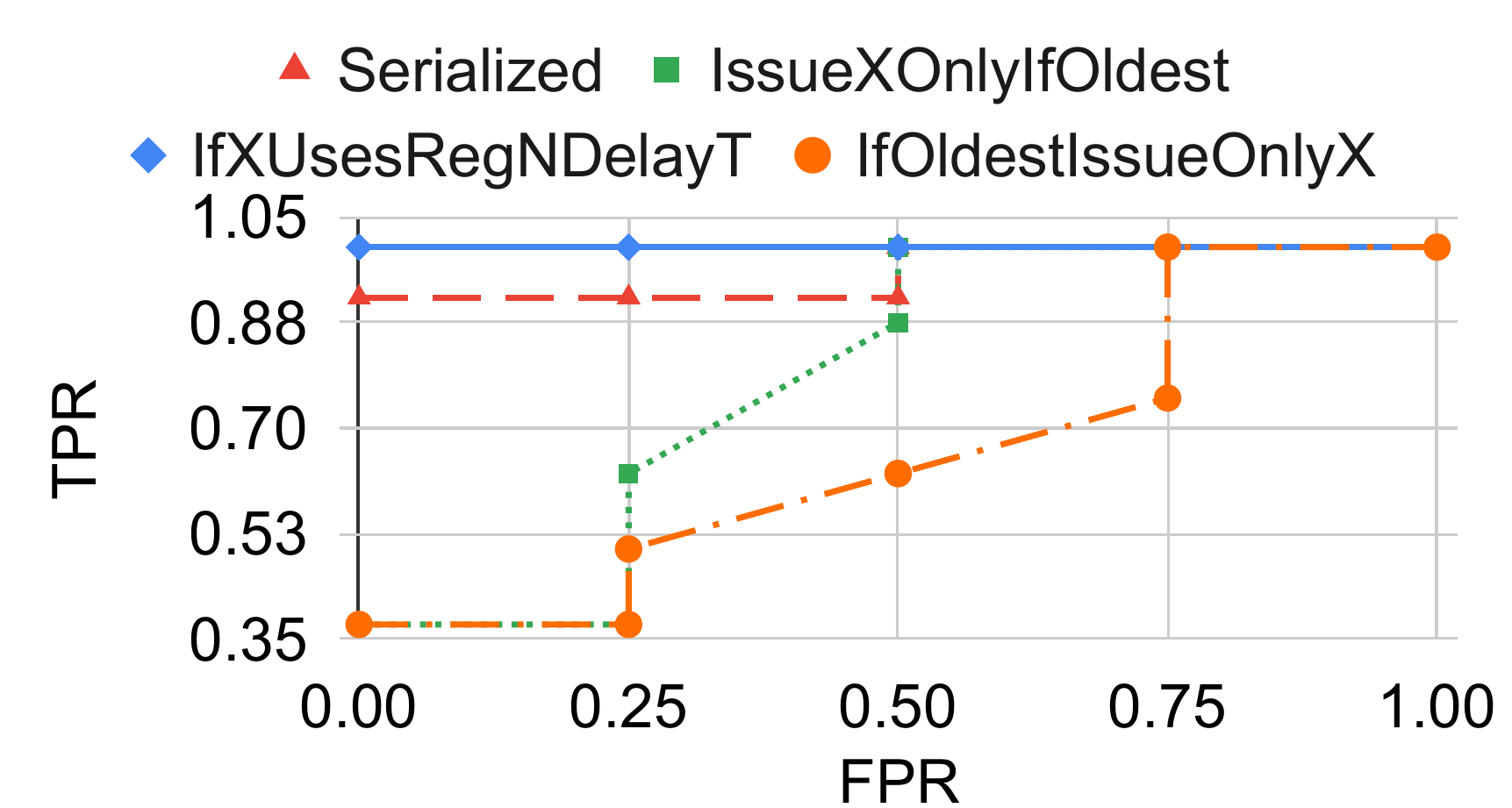}
    \caption{ROC curves for GBT-250 on different bug types.}
    \label{fig:roc_sch_new}
\end{figure}

In Figure \ref{fig:roc_sch_new} we show several examples of how the ROC curve looks for different bug types when stage 1 of our method uses GBT-250 model. Difficult to catch bugs usually have a lower ROC AUC, while other bugs with higher IPC impact can be detected without false-positives.

\subsection{Number of Probes}
\begin{figure}[!hbt]
    \centering
    \includegraphics[width = 0.45\textwidth]{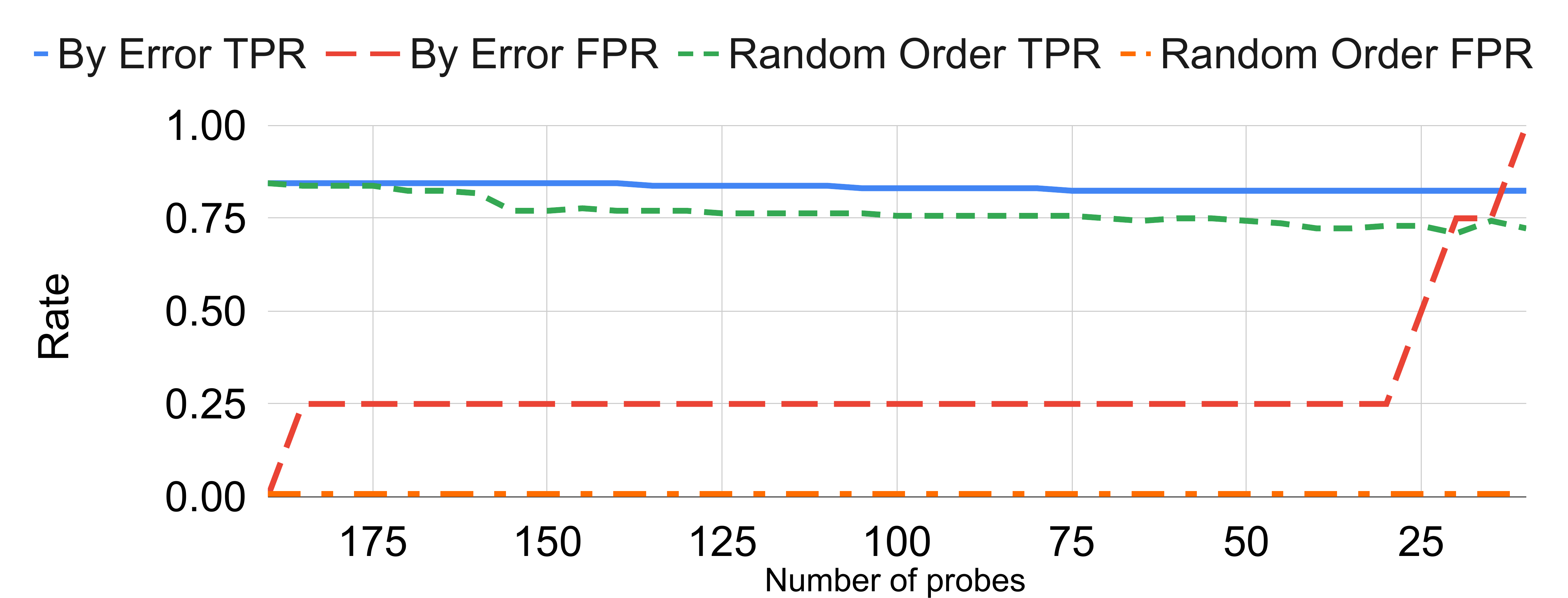}
    \caption{Effect of removing probes.}
    \label{fig:remove_probe}
\end{figure}

There is a trade-off associated with the number of probes. More probes
can potentially detect more bugs or reduce false positives, at the
cost of higher runtime.  To see how accuracy varies with probe count
we examine the impact of reducing the number of probes.
We perform this experiment in an iterative approach. Every iteration
we remove five probes, re-train the model with the reduced set and
collect its detection metrics.  We evaluated these results using the
GBT-250 model.  We perform the probe reduction experiment with
two different orders:

\begin{enumerate}
\itemsep0em 
\item Remove the probes with the highest error in IPC inference
  first. The insight for using this method is, if a probe has high IPC
  inference error, it is likely that the model did not learn from it
  properly.  The results are shown in Figure~\ref{fig:remove_probe}.
\item Remove probes in random order.  This case is equivalent to
  having fewer probes with which to test the design.  Results are
  shown in Figure \ref{fig:remove_probe}.
\end{enumerate}

Results for both orderings, in Figure \ref{fig:remove_probe} show
that, when the number of probes is reduced, the quality of results is
degraded, either by an increase in FPR or by decreasing the TPR.
It is also important to note,
however, that the accuracy change is very slow versus probe reduction.
Thus, our results are quite robust, arguing that even fewer benchmarks
could be used as an input to the process with little impact on the
outcome.

\subsection{Counter Selection}

\begin{figure}[!hbt]
    \centering
    \includegraphics[width = 0.37\textwidth]{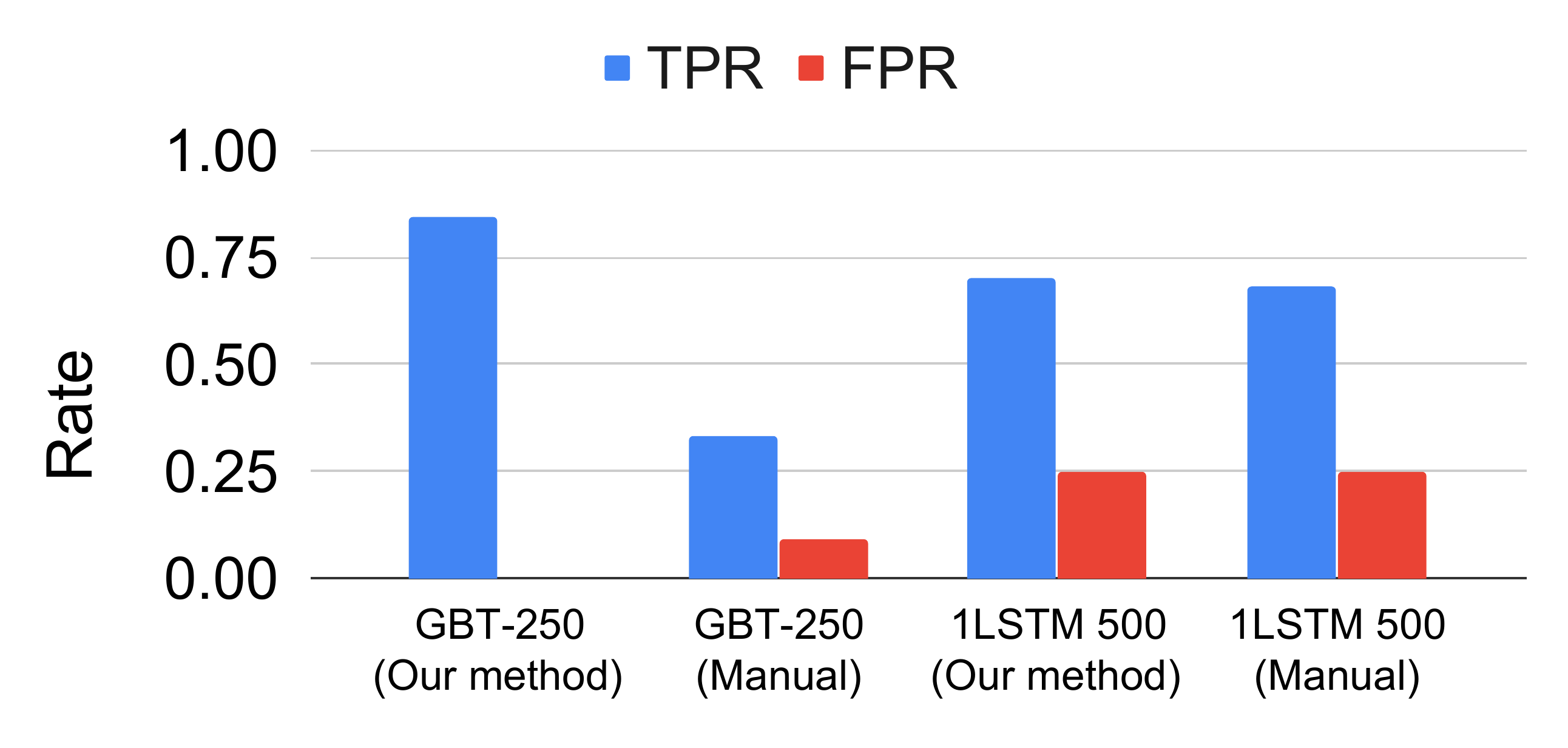}
    \caption{Effect of counter selection method.}
    \label{fig:counter_selection}
\end{figure}

We evaluated our counter selection methodology by comparing with a set
of 22 manually, but not arbitrarily, selected counters, which include
miss rates for different cache levels, branch statistics and other
counters related to the core pipeline and how many instructions each
stage has processed. Unlike our method, we use the same 22 counters
for all the probes.  The results are evaluated for models 1-LSTM-500
and GBT-250.  The obtained results can be seen on Figure
\ref{fig:counter_selection}.

Our counter selection methodology achieves better results in both
machine learning models when compared to the results obtained by
manually selecting the counters. Despite being heuristic, our counter
selection me\-tho\-do\-lo\-gy generally facilitates better detection
results.



\subsection{Time Step Size}

\begin{figure}[!hbt]
    \centering
    \begin{subfigure}[b]{0.23\textwidth}
     \includegraphics[width=0.99\textwidth]{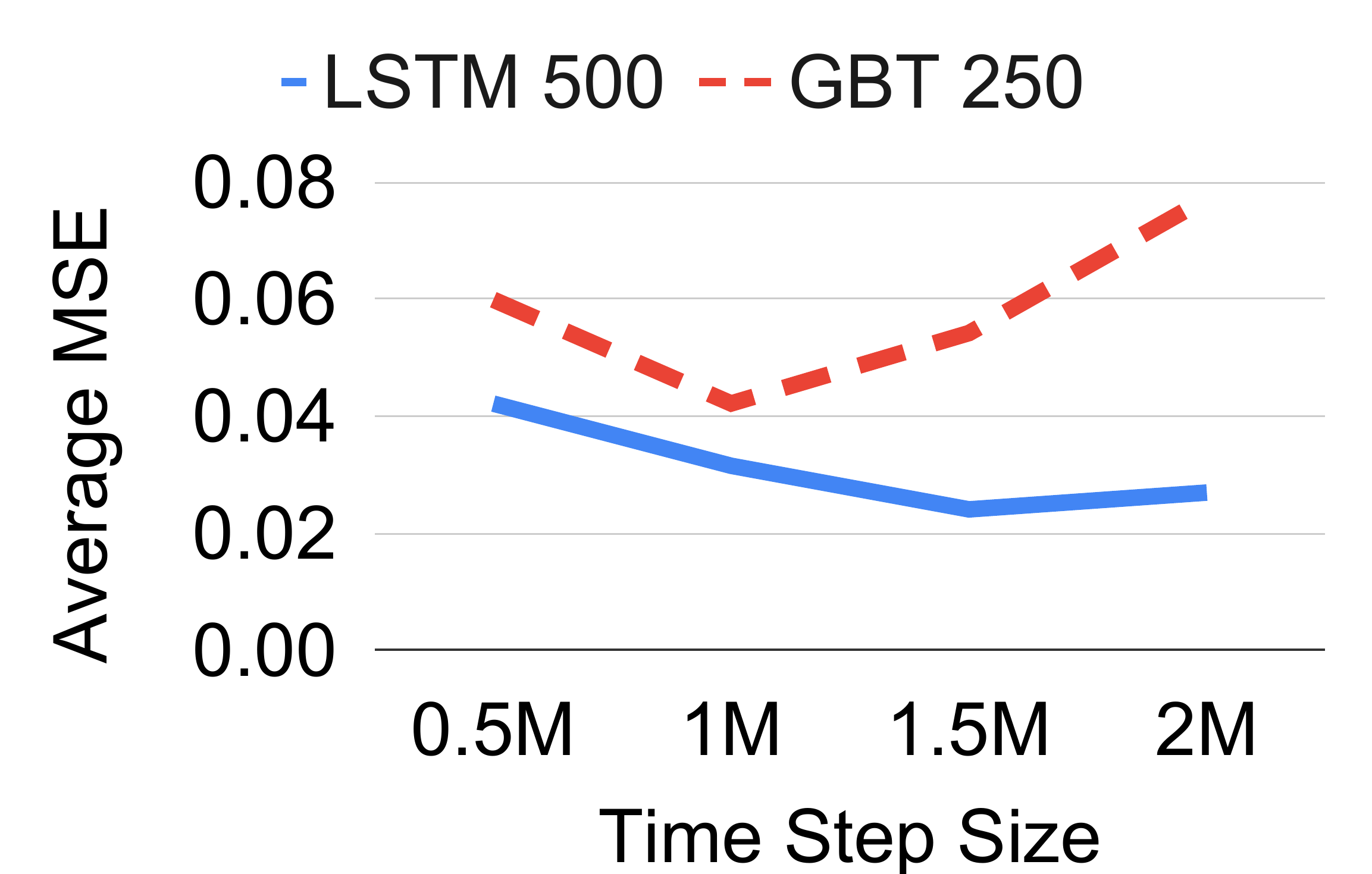}
        \caption{Average MSE across all probes.}
        \label{fig:ts_mse}
    \end{subfigure}
    ~ 
    \begin{subfigure}[b]{0.23\textwidth}
    \includegraphics[width=0.99\textwidth]{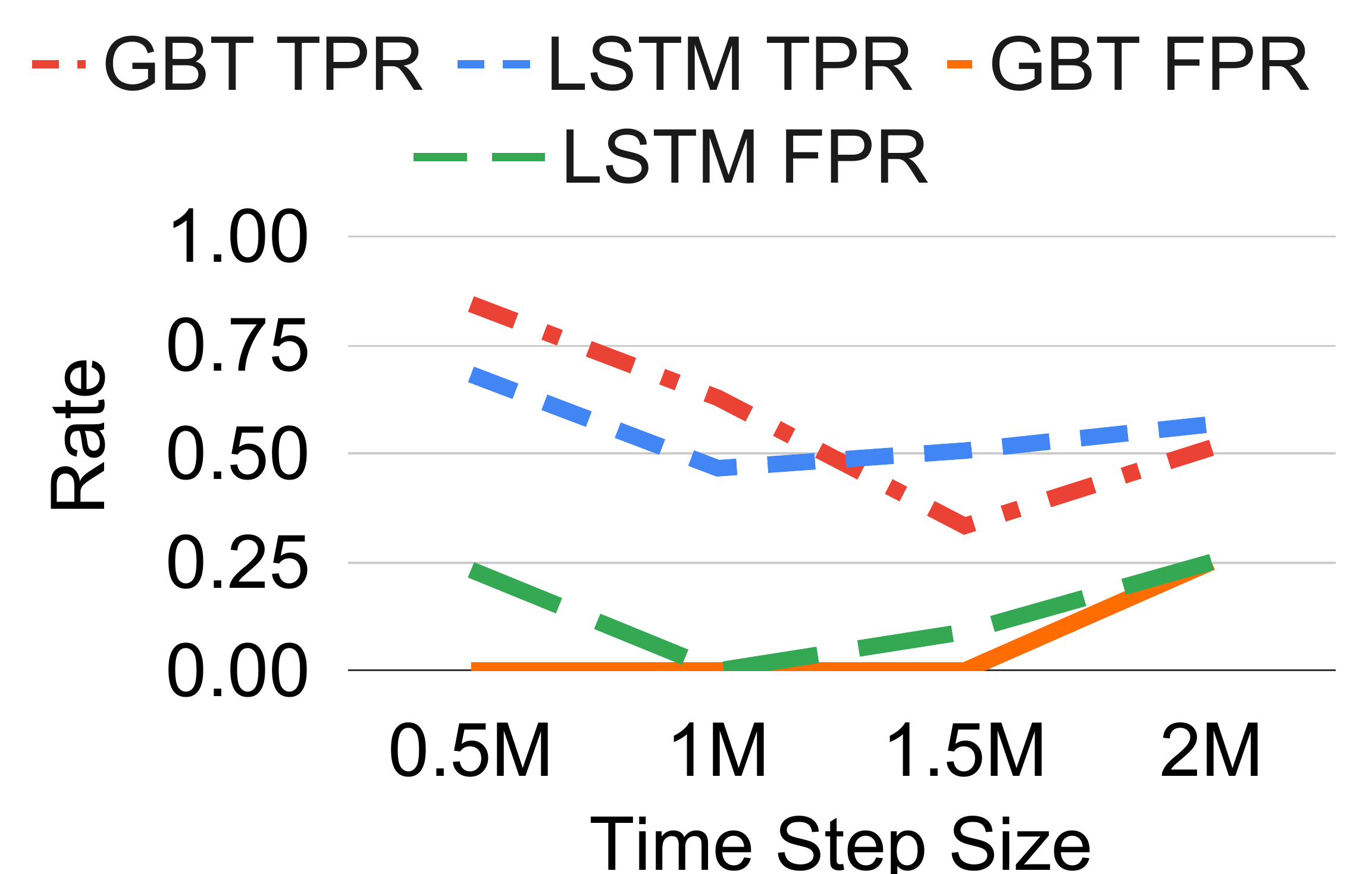}
       \caption{TPR and FPR on bug detection.}
        \label{fig:ts_tpr_fpr}
    \end{subfigure}
    \caption{Effect of different time step sizes.}
    \label{fig:ts}
\end{figure}

In stage 1 of IPC modelling, input features are taken as time series
with each time step being 500K clock cycles. Experiments were
performed to observe the effect of different time step sizes. The
results are plotted in Figure~\ref{fig:ts}.  When the time step size
increases, the inference errors for model 1-LSTM-500 decrease as
shown in Figure~\ref{fig:ts_mse}. This is because coarser grained
inference is generally easier than fine-grained. MSE is
used here instead of the error defined by Eq.~\eqref{eq:error}
as error area among different step sizes series are not
comparable.

The reduced IPC inference errors do not necessarily lead to improved
bug detection results. In fact, Figure~\ref{fig:ts_tpr_fpr} shows that
both TPR and FPR degrade as the time step size increases. The
rationale is that whether or not the IPC inference is sensitive to
bugs matters more than the accuracy.  The results in
Figure~\ref{fig:ts} confirm the choice of 500K cycles as the time step
size.  Besides the efficacy of bug detection, time step size also
considerably affects computing runtime and data storage. In our
experience, step size of 500K cycles reaches a good compromise between
bug detection and computing load in our experiment setting.


\subsection{Window Size}

The IPC inference in stage 1 can take the feature data from a series
of time steps. So far in this paper, the window size we have used in
our experiments has been one because the time step size is
sufficiently large. Experiments were conducted to observe the impact
of increasing the window size. Table \ref{tab:window} shows the TPR
and FPR obtained when the window size is increased for the model
GBT-250.


\begin{table}[!hbt]
\centering
\caption{Window size effect.}
\label{tab:window}
\resizebox{0.23\textwidth}{!}{
\begin{tabular}{c|cccc}
\hline
 & \multicolumn{4}{c}{\textbf{Window Size}} \\
 & 1 & 2 & 3 & 4 \\ \hline
TPR & 0.84 & 0.48 & 0.32 & 0.48 \\
FPR & 0.00 & 0.21 & 0.00 & 0.39 \\ \hline
\end{tabular}
}
\end{table}

The results confirm the choice of a window size of one throughout our
experiments. Given our time step size, the results suggest that adding
information of previous time steps do not help to increase sensibility
to performance bugs, furthermore, it actually degrades it.

\subsection{\hlcyan{Microarchitecture Design Parameter Features}}
\begin{figure}[!hbt]
    \centering
    \includegraphics[width = 0.4\textwidth]{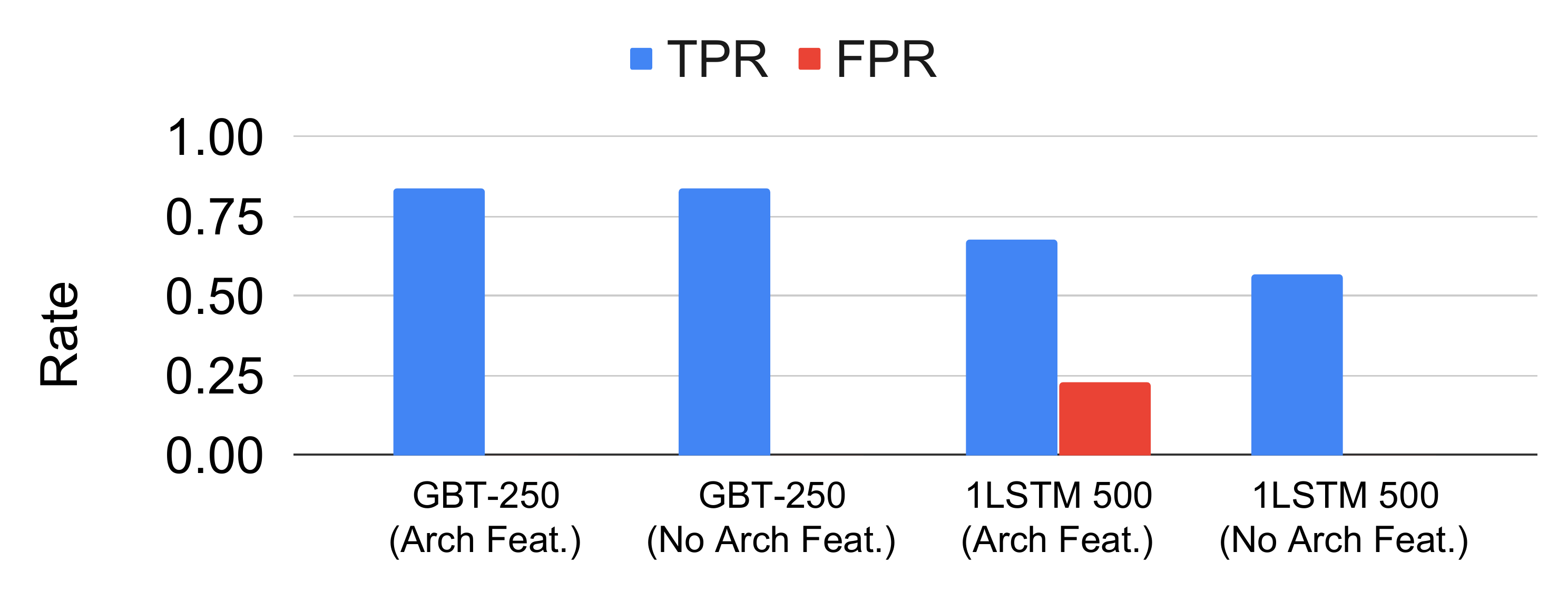}
    \caption{Effect of microarchitecture design parameter features.}
    \label{fig:arch_v_noarch}
\end{figure}

In our baseline methodology we propose the use of microarchitectural
design parameters/specifications as static features (\emph{e.g.}
ROB size, issue width, etc.).  Here, we examine the impact of
removing these static features on the accuracy of our bug detection
methodology. Figure \ref{fig:arch_v_noarch} shows the obtained
results.

The results show removing the design parameters has no impact on the
accuracy of the GBT-250 model and a small reduction on the number of
detected bugs with the 1LSTM-500 model, althought the number of false
alarms is also reduced. This impact is contained within the bugs of
\emph{Low} or \emph{Very Low} IPC impact. These results indicate that
performance impact information is, in many cases, sufficiently
contained within the performance counters (\emph{i.e.} performance
counter data inherently conveys enough information for the model to
infer the IPC of different microarchitectures on the given workloads),
and the change on microarchitectural specifications has a very small
impact on the quality of results for our methodology.

\subsection{Number of training microarchitectures}
\label{subsec_exp_training_uarchs}

\begin{figure}[!hbt]
    \centering
    \includegraphics[width = 0.28\textwidth]{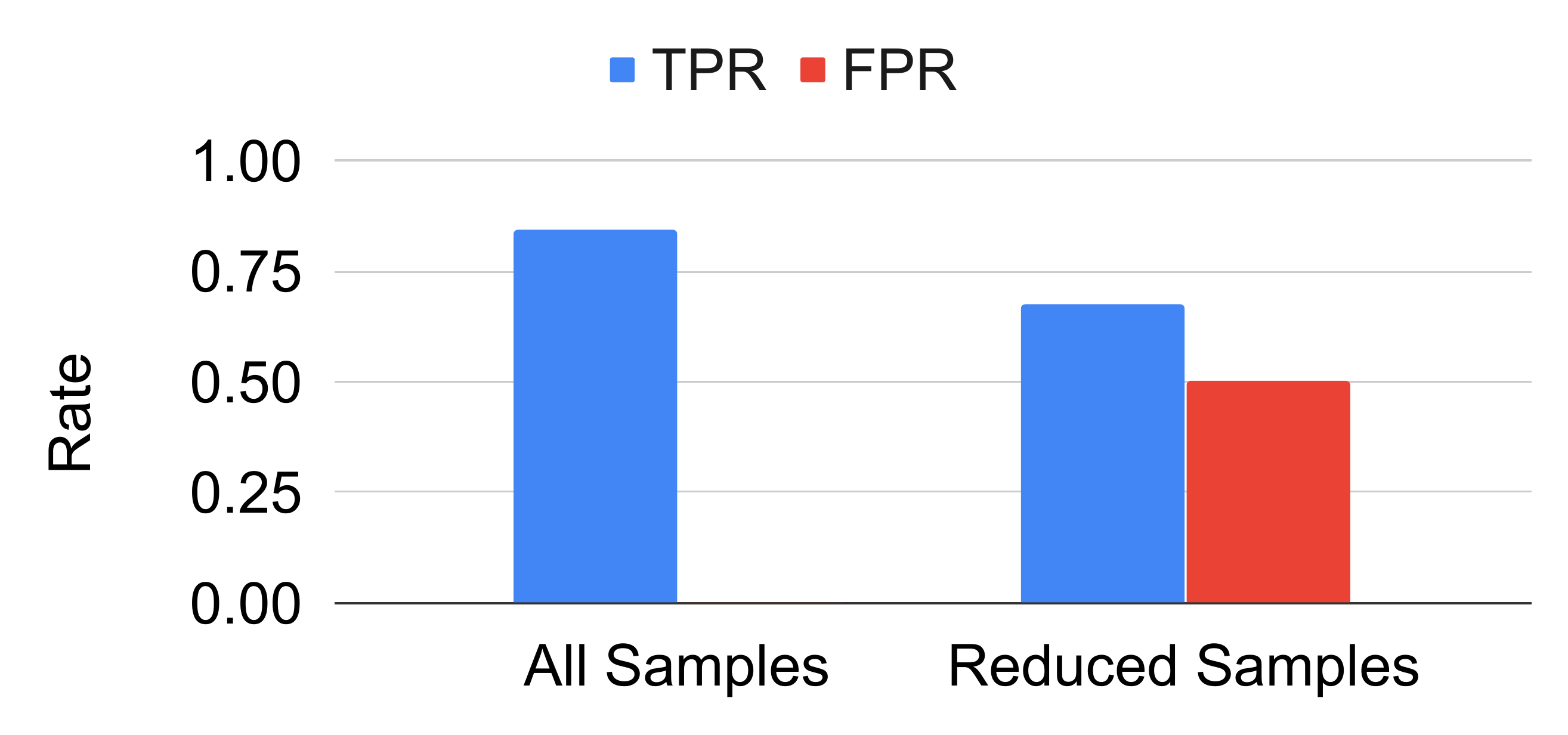}
    \caption{Effect of number of training microarchitectures.}
    \label{fig:less_archs}
\end{figure}

We also evaluated the effect of reducing the number of available
architectures to train our method.  Here, we use 5 microarchitectures
to train our IPC model (Set I), instead of 9. Sets II and III were
reduced from 3 and 4 to 2 and 3 microarchitectures, respectively.  In
each case we dropped the ``artificial'' microarchitectures, keeping
only the real ones.  We keep the number of testing microarchitectures
constant and show the results using GBT-250 in Figure
\ref{fig:less_archs}.

From the obtained results we confirm that creating artificial
architectures is necessary in order to augment our data set. This aids
the model to learn the difference between performance variation due to
microarchitectural specifications and performance bugs.

\subsection{\hlcyan{Bug Detection in Cache Memory Systems}}
\label{subsec_exp_mem}

In this section, we evaluate performance bug detection in the cache
memory system, as described on Section \ref{subsec:detection_mem}. The
obtained results are shown in Table \ref{tab:class_mem}.

\begin{table}[!htb]
\centering
\caption{Bug detection in Memory Systems results.}
\resizebox{0.48\textwidth}{!}{
\label{tab:class_mem}
\begin{tabular}{c|c|c|c|c|cccc}
\hline
 \multicolumn{1}{l|}{} & \multicolumn{1}{l|}{} & \multicolumn{1}{l|}{} & \multicolumn{1}{l|}{} & \multicolumn{1}{l|}{} & \multicolumn{4}{c}{\textbf{TPR for different bug categories}} \\
\textbf{\begin{tabular}[c]{@{}c@{}}Stage 1\\ Metric\end{tabular}} & \textbf{\begin{tabular}[c]{@{}c@{}}Stage 1\\ ML Model\end{tabular}}  & \textbf{FPR} & \textbf{TPR} & \textbf{Precision} & \textbf{High} & \textbf{Medium} & \textbf{Low} & \textbf{Very Low} \\ \hline
\multirow{2}{*}{IPC} & \textbf{LSTM} & 0.00 & 1.00 & 1.00 & 1.00 & 1.00 & 1.00 & 1.00 \\ 
 & \textbf{GBT}  & 0.00 & 1.00 & 1.00 & 1.00 & 1.00 & 1.00 & 1.00 \\
\hline
\multirow{2}{*}{AMAT} & \textbf{LSTM} & 0.00 & 0.82 & 1.00 & 1.00 & 1.00 & 1.00 & 0.20 \\ 
 & \textbf{GBT}  & 0.00 & 1.00 & 1.00 & 1.00 & 1.00 & 1.00 & 1.00 \\
\hline
\end{tabular}
}
\end{table}

These results show that our methodology is able to detect all the bugs
when GBT models are used for both IPC and AMAT inferences, while LSTM
only misses \emph{Very Low} AMAT impact bugs.  The high accuracy of
these results show that this methodology is robust for usage on
different system components beyond the core, as well as different
simulators.

\section{Related Work}
\label{sec:prior}


\subsection{Microprocessor Performance Validation}

The importance as well as the challenge of processor performance debug
was recognized 26 years
ago~\cite{Bose1994,surya1994architectural}. However, the followup
study has been scarce perhaps due to the difficulty. The few known
works~\cite{Bose1994,surya1994architectural,Utamaphethai1999,Singhal2004,Ould2007}
generally follow the same strategy although with different
emphasis. That is, an architecture model or an architecture
performance model is constructed, and compared with the new design on
a set of applications. Then, performance bugs can be detected if a
performance discrepancy is observed in the comparison. In prior work
by Bose~\cite{Bose1994}, functional unit and instruction level models
are developed as golden references. However, a performance bug often
manifests in the interactions among different components or different
instructions. Surya \emph{et al.}~\cite{surya1994architectural} built
an architecture timing model for PowerPC processor. It is focused on
enforcing certain invariants in executing loops. For example, the IPC
for executing a loop should not decrease when the buffer queue size
increases. Although this technique is useful, its effectiveness is
restricted to a few types of performance bugs.
In work by Utamaphethai \emph{et~al.}~\cite{Utamaphethai1999}, a
microarchitecture design is partitioned into buffers, each of which is
modeled by a finite state machine (FSM). Performance verification is
carried out along with the functional verification using the FSM
models. This method is effective only for state dependent performance
bugs. The model comparison-based approach is also applied for the
Intel Pentium 4 processor~\cite{Singhal2004}. Similar approach is also
applied for identifying I/O system performance
anomaly~\cite{shen2005system}. A parametric machine learning model is
developed for architecture performance
evaluation~\cite{Ould2007}. This technique is mostly for guiding
architecture design parameter (e.g., cache size) tuning.

Overall, the model comparison based approach has two main
drawbacks. First, the same performance bug can appear in both the
model and the design as described by Ho \cite{Ho1995}, and
consequently cannot be detected.  Although some works strive to find
an accurate golden reference~\cite{Bose1994}, such effort is
restricted to special cases and very difficult to generalize. In
particular, in presence of intrinsic performance
variability~\cite{cook2017performance,skinner2005understanding},
finding golden reference for general cases becomes almost impossible.
Second, constructing a reference architecture model is labor intensive
and time consuming. On one hand, it is very difficult to build a
general model applicable across different architectures. On the other
hand, building one model for every new architecture is not
cost-effective.

\subsection{Performance Bugs in Other Domains}

A related performance issue in datacenter computing is performance
anomaly detection~\cite{Ibidunmoye2015}.  The main techniques here
include statistics-based, such as ANOVA tests and regression analysis,
and machine learning-based classification. Although there are many
computers in a datacenter, the subject is the overall system
performance upon workloads in very coarse-grained metrics. As such,
the normal system performance is much better defined than individual
processors. Performance bug detection is mentioned for distributed
computing in clouds~\cite{killian2010finding}. In this context, the
overall computing runtime of a task is greater than the sum of
runtimes of computing its individual component as extra time is needed
for the data communication. However, the difference should be limited
and otherwise an anomaly is detected. As such, the performance debug
in distributed computing is focused on the communication and assumes
that all processors are bug-free. Evidently, such assumption does not
hold for processor microarchitecture designs.  Gan \emph{et
  al.}~\cite{gan2019seer} developed an online QoS violation prediction
technique for cloud computing.  This technique applies runtime trace
data to a machine learning model for the prediction and is similar to
our baseline approach in certain extent. In another work, cloud
service anomaly prediction is realized through self-organizing map, an
unsupervised learning technique~\cite{dean2012ubl}.

Performance bug is also studied for software systems where bugs are
detected by users or code reasoning~\cite{Nistor2013}. A machine
learning approach is developed for evaluating software performance
degradation due to code change~\cite{alam2019autoperf}. Software code
analysis~\cite{li2018pcatch} is used to identify performance critical
regions when executing in cloud computing. Parallel program
performance diagnosis is studied by Atachiants \emph{et al.}
\cite{Atachiants2016}. Performance bugs in smartphone applications are
categorized into a set of patterns~\cite{Liu2014} for future
identification. As the degree of concurrency in microprocessor
architectures is usually higher than a software program, performance
debugging for microprocessor is generally much more complicated.



\subsection{Performance Counters for Power Prediction}

\hlcyan{Prior work has used performance counters to predict power
  consumption}~\cite{Joseph2001,Contreras2005,Bircher2007,Yang2016}. \hlcyan{
  Joseph~\emph{et al.}}~\cite{Joseph2001} \hlcyan{aim to predict average
  power consumption on complete workloads, as opposed to the time-series
  based strategy we use for accurate detection of bugs.  Counters are
  selected based on heuristics, without automation. Contreras
  ~\emph{et al.}}~\cite{Contreras2005} \hlcyan{further improves this work
  and present an automated performance counter selection technique
  able to do time-series prediction of the power.
  However, this technique was evaluated on an Intel PXA255 processor,
  a single-issue machine, making the problem much simpler
  than aiming at super-scalar processors.  Bircher \emph{et
    al.}}~\cite{Bircher2007} \hlcyan{further improves this line of work by
  creating models other units, such as memory, disk and
  interconnect. The main drawback of this work is that it requires a
  thorough study of the design characteristics in order to create the
  performance counter list to be used. Recent work by Yang~\emph{et
    al.}}~\cite{Yang2016} \hlcyan{further expanded Bircher's work by
  aiming to develop a full system power model, prior work had only
  focused on component based modeling}.

\hlcyan{Although this line of work has similarities with our IPC
  estimation methodology, our work is the first whose goal is the
  identification of performance bugs in a design.  Further, because
  our goal is performance bug detection via orthogonal probes, we can
  make our models orthogonal and specific to each probe, thus we
  achieve very high IPC estimation accuracy, higher than is possible
  with the generalized models needed for general power
  prediction/management. Another significant difference is that our
  methodology is able to generalize to multiple microarchitectures,
  whereas the methodologies discussed in this section are trained and
  tested on the same processor.  }

\section{Conclusion and Future Work}
\label{sec:conc}

In this work, a machine learning-based approach to automatic
performance bug detection is developed for processor core
microarchitecture designs. The machine learning models extract
knowledge from legacy designs and avoid the previous methods of
reference performance models, which are error prone and time consuming
to construct.  Simulation results show that our methodology can detect
91.5\% of the bugs with impact greater than 1\% of the IPC on a new
microarchitecture when completely new bugs exist in a new
microarchitecture. With this study we also hope to draw the attention
of the research community to the broader performance validation
domain.

In future research, we will extend the methodology for debugging of
multi-core memory systems and on-chip communication
fabrics. \hlcyan{We will also further study how to automatically
  narrow down bug locations once they are detected.  As in functional
  debug}~\cite{BugMD}, \hlcyan{ the results from our method will
  potentially serve as symptoms for bug localization.  By analyzing
  characteristics in common across the probes triggering the bug
  detection (\emph{e.g.} most common instruction types, memory or
  computational boundness, etc), the designers could reduce the list
  of potential bug locations. Another debugging path could be the
  analysis of the counters selected for the IPC inference models of
  those traces. Factors such as a lost of correlation between the
  counters and IPC when compared to legacy designs could also help to
  pinpoint possible sources of bug.}


\section*{Acknowledgements}
This work is partially supported by Semiconductor Research Corporation Task 2902.001.


\bibliographystyle{IEEEtranS}
\bibliography{refs}

\end{document}